\documentclass[8.5pt,twoside,twocolumn]{article}
\usepackage[super,sort&compress,comma]{natbib} 
\usepackage{mhchem}
\usepackage{times,mathptmx}
\usepackage{sectsty}
\usepackage{balance}

\usepackage{graphicx} %eps figures can be used instead
\usepackage{lastpage}
\usepackage[format=plain,justification=raggedright,singlelinecheck=false,font=small,labelfont=bf,labelsep=space]{caption} 
\usepackage{fancyhdr}
\usepackage{amsmath}
\usepackage{mathrsfs}
\usepackage{amssymb}
\newcommand{\bm}[1]{\boldsymbol{#1}}

\begin{document}

\thispagestyle{plain}
\fancypagestyle{plain}{
%\fancyhead[L]{\includegraphics[height=8pt]{headers/LH}}
%\fancyhead[L]{Optical antennas as nanoscale resonators}
%\fancyhead[C]{\hspace{-1cm}\includegraphics[height=20pt]{headers/CH}}
%\fancyhead[R]{\includegraphics[height=10pt]{headers/RH}\vspace{-0.2cm}}
\renewcommand{\headrulewidth}{1pt}}
\renewcommand{\thefootnote}{\fnsymbol{footnote}}
\renewcommand\footnoterule{\vspace*{1pt}% 
\hrule width 3.4in height 0.4pt \vspace*{5pt}} 
\setcounter{secnumdepth}{5}

\makeatletter 
\def\subsubsection{\@startsection{subsubsection}{3}{10pt}{-1.25ex plus -1ex minus -.1ex}{0ex plus 0ex}{\normalsize\bf}} 
\def\paragraph{\@startsection{paragraph}{4}{10pt}{-1.25ex plus -1ex minus -.1ex}
{0ex plus 0ex}{\normalsize\textit}} 
\renewcommand\@biblabel[1]{#1}            
\renewcommand\@makefntext[1]% 
{\noindent\makebox[0pt][r]{\@thefnmark\,}#1}
\makeatother 
\renewcommand{\figurename}{\small{Fig.}~}
\sectionfont{\large}
\subsectionfont{\normalsize} 

\fancyfoot{}
%\fancyfoot[LO,RE]{\vspace{-7pt}\includegraphics[height=9pt]{headers/LF}}
%\fancyfoot[CO]{\vspace{-7.2pt}\hspace{12.2cm}\includegraphics{headers/RF}}
%\fancyfoot[CE]{\vspace{-7.5pt}\hspace{-13.5cm}\includegraphics{headers/RF}}
%\fancyfoot[RO]{\footnotesize{\sffamily{1--\pageref{LastPage} ~\textbar  \hspace{2pt}\thepage}}}
%\fancyfoot[LE]{\footnotesize{\sffamily{\thepage~\textbar\hspace{3.45cm} 1--\pageref{LastPage}}}}
\fancyfoot[RO]{\footnotesize{\sffamily{\thepage}}}
\fancyfoot[LE]{\footnotesize{\sffamily{\thepage}}}
\fancyhead{}
\renewcommand{\headrulewidth}{1pt} 
\renewcommand{\footrulewidth}{1pt}
\setlength{\arrayrulewidth}{1pt}
\setlength{\columnsep}{6.5mm}
\setlength\bibsep{1pt}

\twocolumn[
  \begin{@twocolumnfalse}
\noindent\LARGE{\textbf{Optical antennas as nanoscale resonators}}
\vspace{0.6cm}

\noindent\large{\textbf{Mario Agio,$^{\ast}$\textit{$^{a}$}}}
\vspace{0.5cm}
%Please note that \ast indicates the corresponding author(s) but no footnote text is required.

%\noindent\textit{\small{\textbf{Received Xth XXXXXXXXXX 20XX, Accepted Xth XXXXXXXXX 20XX\newline First published on the web Xth XXXXXXXXXX 200X}}}

%\noindent \textbf{\small{DOI: 10.1039/b000000x}}
%\vspace{0.6cm}
%Please do not change this text.

\noindent \normalsize{
Recent progress in nanotechnology has enabled us to fabricate subwavelength
architectures that function as antennas for improving the 
exchange of optical energy with nanoscale matter.
We describe the main features of optical antennas for
enhancing quantum emitters
and review designs that increase the spontaneous emission rate by
orders of magnitude from the ultraviolet up to the near-infrared spectral range. 
To further explore how optical antennas may lead
to unprecedented regimes of light-matter interaction,
we draw a relationship between metal
nanoparticles, radio-wave antennas and optical resonators.
Our analysis points out how optical antennas may function as
nanoscale resonators and how these may offer unique
opportunities with respect to state-of-the-art microcavities.}
\vspace{0.5cm}
 \end{@twocolumnfalse}
  ]

\section{Introduction}

\footnotetext{\textit{$^{a}$~ETH Zurich, Laboratory of Physical Chemistry,
Wolfgang-Pauli-Str. 18, 8093 Zurich, Switzerland. Fax: +41 44 633 1316; Tel: +41
 44 632 3322; E-mail: mario.agio@phys.chem.ethz.ch}}
%\footnotetext{\textit{$^{b}$~Address, Address, Town, Country. }}

The dramatic advances of nanotechnology experienced in recent years
have fueled much interest in optical antennas as
devices for managing the concentration, absorption and radiation of
light at the nanometer scale.\cite{grober97,pohl00,muehlschlegel05,schuck05} In fact,
the amount of activities on this topic has grown very rapidly in various
fields of research, spanning physics, chemistry, electrical engineering,
biology, and medicine to cite a
few.\cite{bharadwaj09,anker08,schuller10,novotny11,biagioni11,giannini11,mayer11}
At a more fundamental level, these systems may
enhance the radiation properties of
quantum emitters, such as atoms and molecules,\cite{greffet05}
an endeavor that dates back to the onset of field-enhanced
spectroscopy.\cite{drexhage74,chance78,metiu84,moskovits85}

Somewhat in parallel, the past decades have witnessed
great progress in the fundamentals and applications
of optical resonators.\cite{haroche89,vahala03,benisty99}
In particular, recent developments in photonic crystals have
enabled the realization of miniaturized cavities with mode
volumes of the order of one cubic wavelength and huge
quality factors.\cite{akahane03,song05}
Obtaining resonators with even smaller dimensions is
a current research challenge, which pushes optical physics
and nanofabrication into new pathways.

A promising strategy relies on metal nanocavities, which
use metal mirrors to confine light into tight volumes.
They are being explored, for instance, to realize ultrasmall
lasers,\cite{hill07,oulton09,zhu11} and to enhance the spontaneous
emission (SE) rate of quantum emitters.\cite{kroekenstoel09,maksymov10,bulu11}
As resonators are pushed towards deep subwavelength
dimensions, it becomes apparent that their differences
with respect to optical antennas begin to vanish.
In fact, several phenomena and functionalities are investigated
using antenna architectures treated as nanoscale
cavities.\cite{bergman03,protsenko05,noginov09,stockman10,savasta10,ridolfo10}

To gain insight on this exciting scenario for light-matter interaction,
we attempt to uncover the relationship between
optical antennas and resonators.
First, we review a number of empirical rules
to engineer optical antennas that lead to a strong enhancement
of the SE rate with minimal losses
caused by absorption in real metals.\citep{rogobete07a} 
Moreover, we describe designs that are fully compatible with
state-of-the-art nanofabrication and highlight effects related to the
antenna composition and
shape.\cite{rogobete07a,agio07,mohammadi08b,mohammadi09a,mohammadi10}

Next, we consider a simplified antenna model and discuss
basic properties starting from analytical expressions. 
Since the physical dimensions are smaller than the operating
wavelength, we base our analysis on the fundamental
limitations of electrically small antennas.\cite{hansen81}
We thus select a few popular resonator
designs\cite{vahala03,song05} and compare their figures
of merit with those of optical antennas.\cite{mohammadi08b}
We show that the enhancement of
light-matter interaction is comparable to 
that achievable with state-of-the-art cavities.
Therefore, despite absorption by real metals, there is a 
window of opportunity where optical antennas may function as nanoscale 
resonators with a tiny device footprint and manageable losses. 

There is another important advantage. Having
a low quality factor, optical antennas are fully compatible with methods and
techniques of ultrafast spectroscopy\cite{zewail00} and
coherent control.\cite{rabitz00}
We will return to these aspects in the conclusions.

\section{Optical antennas}
\label{optical-antennas:chapter}

\begin{figure}[!htb]
\centering{
\includegraphics[width=8.25cm]{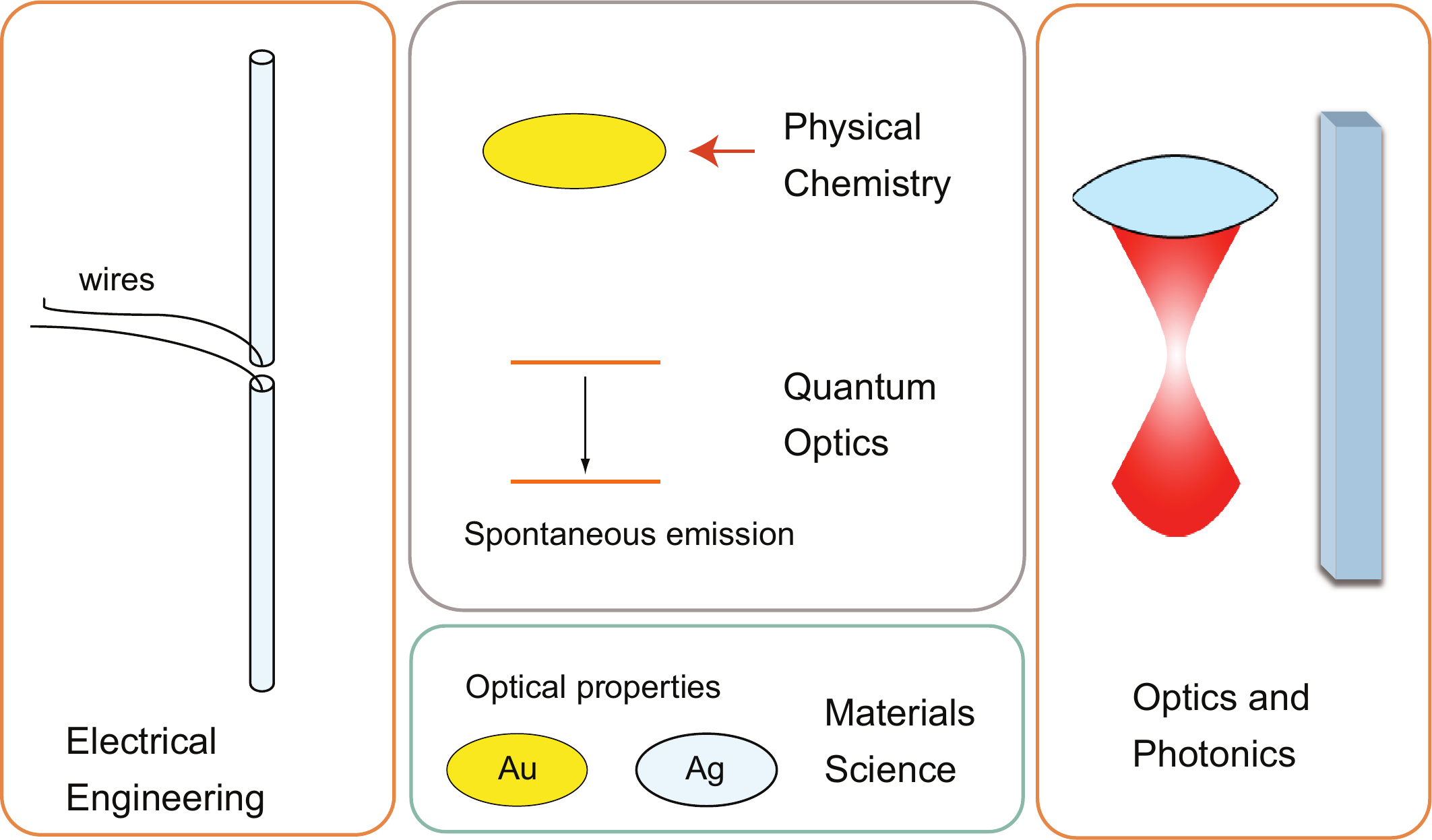}}
\caption{\label{optical-antennas:optical-antennas}
Optical antennas: a truly interdisciplinary research field that
involves diverse areas like electrical engineering, physical chemistry,
quantum optics, materials science, as well as optics and photonics.}
\end{figure}

Optical antennas are metal nanostructures that convert strongly 
localized energy into radiation and vice versa with a high 
throughput.\cite{grober97} 
They share several concepts of radio-wave antennas,
but they also have distinctive features, which are illustrated in
Fig.~\ref{optical-antennas:optical-antennas}. First, the coupling
between the antenna and its load is not via wired electric currents,
but via displacement currents proportional to the near field,
which makes the interaction strongly position and
polarization dependent.\cite{gersten80}
Second, the load is typically a quantum
system, like an atom or a molecule, and as such it is affected by
quantum electrodynamics (QED) phenomena associated with the
modification of the local
electromagnetic environment.\cite{gersten81,ruppin82} 
Third, metals at optical frequencies
are not perfect conductors and their optical properties are strongly
affected by the existence of surface plasmon-polariton (SPP)
resonances.\cite{gersten80,bohren83b}
These modes are tightly
confined and can be controlled at the nanoscale by shaping metals
using state-of-the-art nanofabrication.\cite{barnes03}
Furthermore, they also
depend on intrinsic material properties such as the optical
constants\cite{johnson72} and the electron mean free path.\cite{kreibig08}
Fourth, in optics we
often work with focused beams and guided waves. These should be
considered as relevant degrees of freedom for interfacing light
with optical antennas.\cite{mojarad08,mojarad09a,chen09,chen10} 
In summary, optical antennas represent a
truly interdisciplinary effort that involves electrical engineering,
physical chemistry, quantum optics, materials science as well as
optics and photonics. In this respect, there are ongoing efforts
aimed at their understanding and modeling
within the established and powerful formalism
developed for radio-wave antennas. 
These include the definition of
antenna resonance wavelength\cite{novotny07}
and impedance.\citep{alu08,greffet10}

In Sec.~\ref{optical-antennas:fluorescence}-\ref{optical-antennas:shape}
we explain how optical antennas may enhance light-matter interaction at
the level of a single quantum emitter and how this can be optimized by
design. In Sec.~\ref{optical-antennas:material} we analyze
the main effects associated with the antenna composition and
background medium. Moreover, in Sec. \ref{nanoscale-resonators:towards}
we make use of radio-wave antenna theory to formulate a link
with nanoscale resonators. We thus plan to cover
most of the aspects illustrated in
Fig.~\ref{optical-antennas:optical-antennas}, hoping to show
how an interdisciplinary approach may facilitate our understanding and
also reveal the exciting opportunities of this vibrant research field.

\subsection{Enhancement and quenching of fluorescence}
\label{optical-antennas:fluorescence}

We review the basic phenomena that take place when 
a quantum emitter interacts with a metal nanostructure and try
to make a connection with concepts familiar to radio-wave antennas.
We limit our analysis to the weak excitation limit, where the semi-classical
theory of light-matter interaction is greatly simplified.\cite{meystre07}
The relevant quantities that need to be considered when an emitter
is coupled to an optical antenna are
the field enhancement, the SE rate,
the quantum yield and the radiation pattern.
The last topic does not fall in the focus of this work and will not
be addressed.\cite{li07,hofmann07,kuehn08,taminiau08a,kosako10,curto10}

Under weak resonant excitation
the fluorescence signal can be approximated by the formula
\begin{equation}
S_o=\xi_o\eta_o|\mathbf{E}_o\cdot\mathbf{d}|^2.
\end{equation}
The parameter $\xi_o$ represents the collection
efficiency, $\mathbf{d}$ is the transition electric 
dipole moment, and $\mathbf{E}_o$ is the electric field at the emitter
position. $\eta_o=\Gamma_\mathrm{r}^o/\Gamma_\mathrm{t}^o$
is the quantum yield and it corresponds
to the ratio between the radiative
and total decay rates.
The latter takes into account
the fact that the excited state can also lose energy via
non-radiative channels, i.e. $\Gamma_\mathrm{t}^o=\Gamma_\mathrm{r}^o+
\Gamma_\mathrm{nr}^o$. The label $o$ indicates that these
quantities refer to an isolated emitter.

\subsubsection{Field enhancement.~~}

Away from saturation the excitation rate may be increased by placing
the emitter near a nanostructure that modifies the electric field.
Engineering textbooks do not discuss the intensity enhancement $K$,
because it is not an important design parameter for radio-wave
antennas,\cite{balanis05} while in optical domain the phenomenon has been
thoroughly investigated in the context of surface-enhanced
Raman spectroscopy.\cite{metiu84,moskovits85}
Pioneering works based on polarizability models indicated the
SPP resonance and the lighting rod effect as the 
two most important electromagnetic
enhancement mechanisms.\cite{gersten80} The latter can
be intuitively explained by considering the increase in the surface
charge density $\sigma$ with the curvature of a metal surface.\cite{jackson99}
Since the near field is directly proportional to $\sigma$, nanoparticles
with sharp tips tend to exhibit larger enhancements than nanospheres.
Other strategies to improve the strength of the near field include
the exploitation of nanoscale gaps between two nanoparticles,\cite{aravind81}
the suppression of radiative broadening\cite{wokaun82} and the
choice of different metals.\cite{cline86,zeman87}
These basic design concepts have been applied with more breadth
and detail in the subsequent years, when computational methods
for nano-optics have become available.\cite{girard96}

\subsubsection{Decay rates.~~}

It is well known that the SE rate is not
an intrinsic property of an atom or a molecule, but it also depends
on the local electromagnetic environment.\cite{purcell46} 
Its modification can be obtained by computing
the power emitted by a classical dipole placed in proximity of
the optical antenna. 
The correspondence between quantum and classical theory is valid if
the normalized quantities are used,\cite{chance78,wylie84,xu00}
\begin{equation}
\label{optical-antennas:decay-rad}
\frac{\Gamma_\mathrm{r}}{\Gamma_\mathrm{r}^o}=\frac{P_\mathrm{r}}{P_o},
\end{equation}
where $P_o$ and $P_\mathrm{r}$ are the power radiated by a classical
dipole in free space and near an optical antenna, respectively.

We take advantage of the reciprocity
argument\cite{balanis05} to state that a strong $K$
is associated with a strong modification of the radiative
decay rate. Indeed, it can be shown that for an antenna
that does not modify the radiation pattern of the emitter these
two quantities are exactly equal.\cite{taminiau08b}
Therefore, one could simply refer to the design strategies
discussed in the previous section to obtain a large modification
of the SE rate.\cite{blanco04}

Because part of the emitted power is absorbed by metal losses,
a full characterization of the system requires the calculation of both
radiative $\Gamma_\mathrm{r}$ and non-radiative $\Gamma_\mathrm{nr}$
decay rates.\cite{chance78,kaminski07} The total decay rate 
$\Gamma_\mathrm{t}$ is thus
$\Gamma_\mathrm{r}+\Gamma_\mathrm{nr}+\Gamma_\mathrm{nr}^o$.
The corresponding classical quantities are
easily derived from Poynting theorem,\cite{jackson99} which leads to
\begin{equation}
\label{optical-antennas:decay-tot}
\frac{\Gamma_\mathrm{r}+\Gamma_\mathrm{nr}}{\Gamma_\mathrm{r}^o}=
\frac{P_\mathrm{t}}{P_o},
\end{equation}
where $P_\mathrm{t}$ is the total power dissipated by the dipole.

\subsubsection{Antenna efficiency.~~}

The enhancement of $\Gamma_\mathrm{nr}$ requires some attention.
We follow an approach borrowed from antenna theory,\cite{balanis05} where
the antenna efficiency $\eta_\mathrm{a}$ is defined as the ratio
between the radiated power and the total power
transferred from the load to the antenna.
For the case of a {\it quantum} load, i.e. an
atom or a molecule, it reads
$\eta_\mathrm{a}=\Gamma_\mathrm{r}/(\Gamma_\mathrm{r}+\Gamma_\mathrm{nr})$
and the modified quantum yield $\eta$ takes the expression\cite{lakowicz05}
\begin{equation}
\label{optical-antennas:apparent-quantum-yield}
\eta=\dfrac{\eta_o}{(1-\eta_o)\Gamma_\mathrm{r}^o/\Gamma_\mathrm{r}+
\eta_o/\eta_\mathrm{a}}.
\end{equation}

In comparison with the field enhancement, in the past years
less attention has been dedicated to the improvement of
$\eta_\mathrm{a}$. It turns out that
the latter is mostly affected by higher-order SPP modes,
which are strongly damped by absorption.\citep{gersten81,nitzan81,ruppin82} 
In fact, $\Gamma_\mathrm{nr}$ takes over $\Gamma_\mathrm{r}$
as the emitter approaches the metal surface, because the
source field becomes so inhomogeneous across
the antenna that multipoles are excited more efficiently.
Moreover, there is a contrast between $K$ and $\eta_\mathrm{a}$.
For example, while radiative effects reduce the
near-field strength,\cite{wokaun82} they tend to increase
$\eta_\mathrm{a}$.\cite{mertens07}

Obtaining a large increase of the
SE rate without compromising $\eta_\mathrm{a}$
is thus a non trivial task.
In what follows we show that this is not a fundamental limitation
and discuss situations where the SE rate is
enhanced by more than three orders of magnitude without quenching. 

\subsection{Design rules}

The key design principles for achieving a strong modification
of the SE rate with minimal suffering from
$\Gamma_\mathrm{nr}$ can be summarized as follows.
First, tailor the geometry such that the SPP resonance of
the antenna lies in a spectral region that minimizes
dissipation in the metal. Second, choose elongated 
objects to benefit from strong near fields at sharp corners.
Third, adjust the emitter orientation 
such that its electric dipole moment is aligned with 
that of the antenna.
Fourth, ensure that in the antenna higher order SPP modes
are spectrally separated from the dipolar one.\cite{rogobete07a}
Fifth, choose the antenna volume such that radiation is stronger
than absorption.\cite{mertens07,mohammadi09b} 

\begin{figure*}[!hbt]
\centering{
\includegraphics[width=14cm]{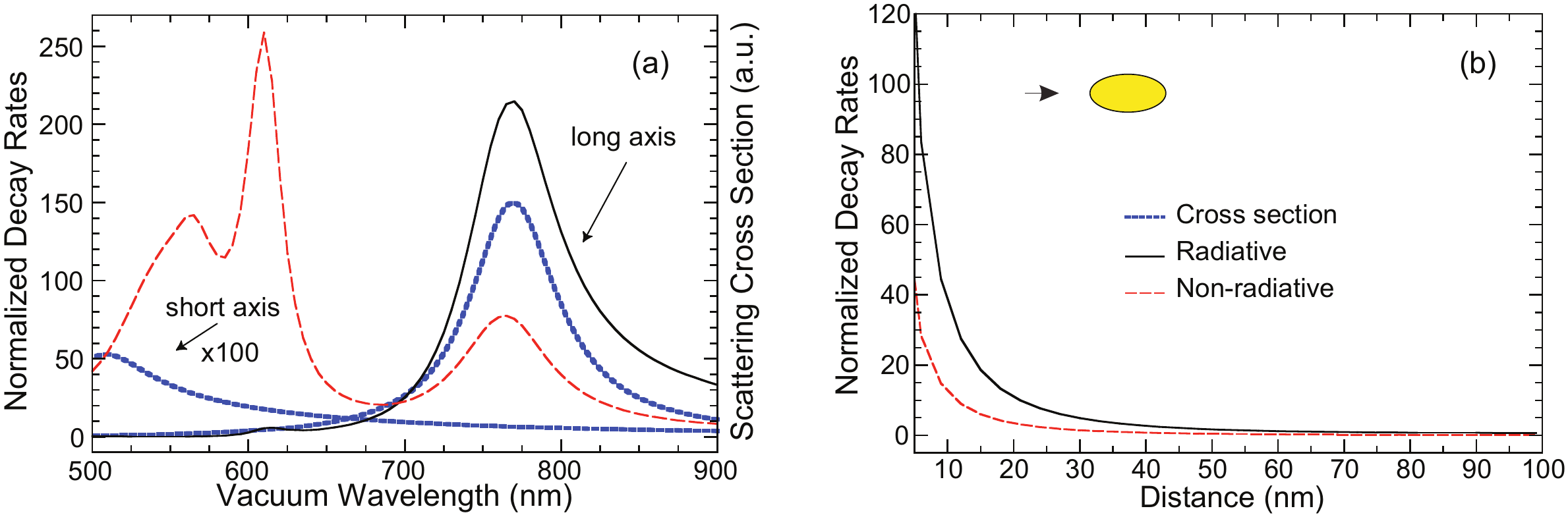}}
\caption{\label{optical-antennas:design-antenna-efficiency}
Normalized decay rates for an emitter coupled to a gold ellipse
(long axis = 60 nm, short axis = 10 nm, backround refractive
index $n_\mathrm{b}$=1.7, two-dimensional model).
(a) Wavelength dependence for a particle-emitter
distance of 3 nm. The scattering cross section of the antenna
is also plotted.
(b) Distance dependence for the wavelength $\lambda$ = 770 nm.
The emitter is oriented as shown in the inset.
Figure adapted with permission from Ref.~\cite{rogobete07a}.
Copyright (2007) by the Optical Society of America.}
\end{figure*}

To exemplify these rules we consider the emission
of a dipole close to an elliptical gold nanoparticle. 
Its SPP resonance is located in the near-infrared region,
where the imaginary part of 
the dielectric function of gold is smaller.\cite{CRChandbook}
Moreover, $K$ is expected to be stronger at the nanoparticle apex.
As shown in Fig.~\ref{optical-antennas:design-antenna-efficiency}a, we see 
that, although both $\Gamma_\mathrm{􏰫r}$ and $\Gamma_\mathrm{nr}$
experience a considerable enhancement, $\Gamma_\mathrm{􏰫r}$
is larger than $\Gamma_\mathrm{nr}$ at the SPP resonance of the long axis. 
Figure~\ref{optical-antennas:design-antenna-efficiency}b
plots the distance dependence of the decay rates at
the long-axis SPP resonance, illustrating
that $\Gamma_\mathrm{􏰫r}$ dominates for all separations larger than 3 nm. 
The strong quenching observed at shorter
wavelengths is attributed to the excitation of higher-order multipoles,
which are spectrally separated from the SPP dipole mode.\cite{rogobete07a}

\subsection{Shape dependence}
\label{optical-antennas:shape}

In Fig.~\ref{optical-antennas:design-antenna-efficiency}
we have shown that changing the shape of the optical antenna can
have a huge impact on its performances.
In this section we analyze this in more detail, with emphasis on
the modification of the SE rate and $\eta_\mathrm{a}$.
In particular, we pay attention to systems and parameters
that are within the reach of standard nanofabrication methods
and of the common experimental techniques used in
nano-optics.\cite{biagioni11}

\subsubsection{Adding a second nanoparticle.~~}

\begin{figure}[h!]
\centering{
\includegraphics[width=5cm]{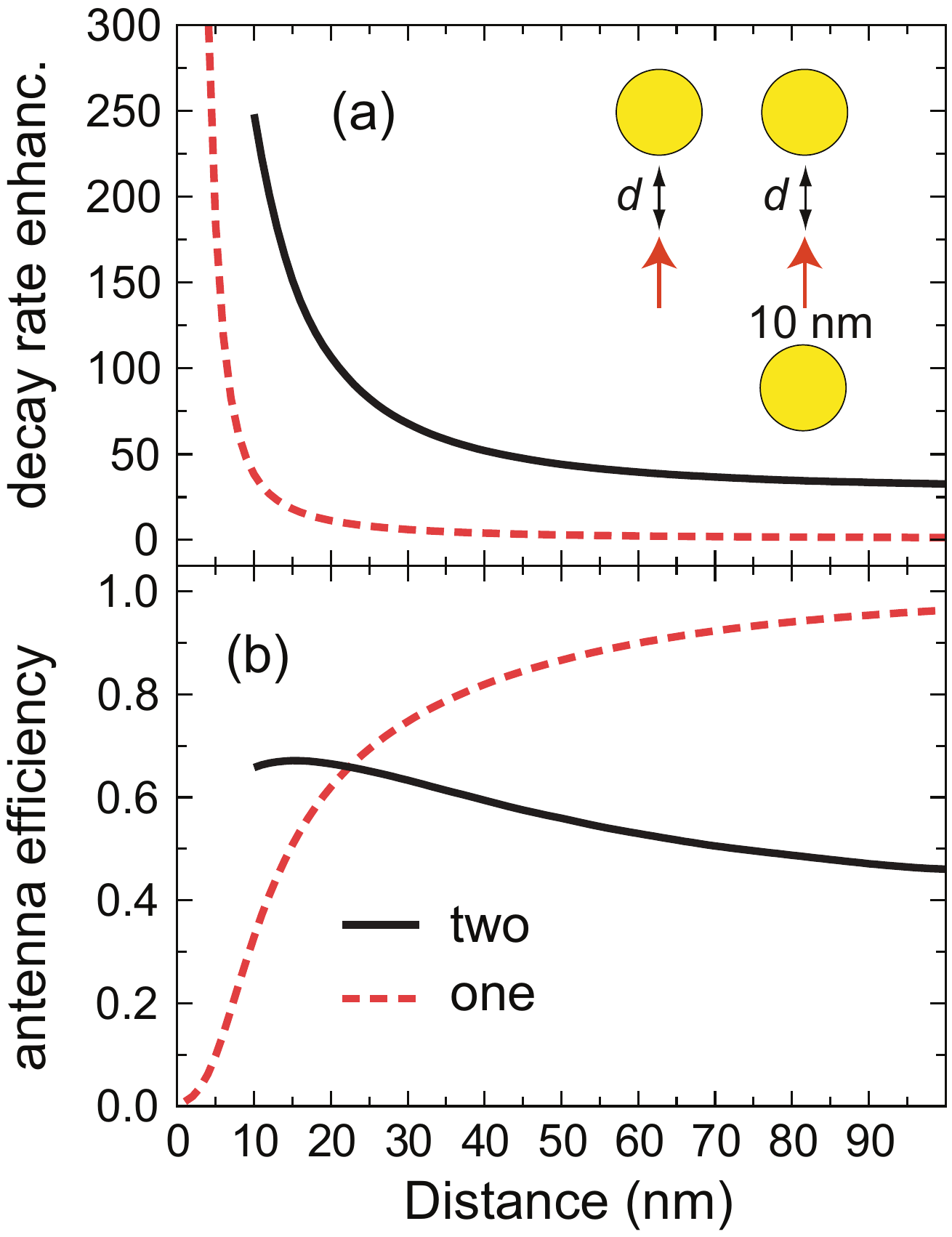}}
\caption{\label{optical-antennas:antenna-design}
Emitter coupled to one or two 100 nm gold nanospheres in air.
Enhancement of the SE rate (a) and
$\eta_\mathrm{a}$ (b) at $\lambda=580$ nm.
The inset describes the coupling scheme. 
Figure adapted with permission from Ref.~\cite{agio07}.
Copyright (2007) by SPIE.}
\end{figure}

A better comparison with the single nanoparticle case can be seen if the molecule is
at a fixed distance from one of the two nano-objects, while the other one is
approached from far away.\cite{agio07} 
The inset of Fig.~\ref{optical-antennas:antenna-design} schematically
shows how the coupling between emitter and antenna is modified by 
changing the distance $d$. 

The enhancement of the SE rate is plotted in
Fig.~\ref{optical-antennas:antenna-design}a 
as a function of $d$.
When both nanoparticles are close to the
emitter, the increase is clearly larger than for a single one.
Figure~\ref{optical-antennas:antenna-design}b shows that
for one nanoparticle $\eta_\mathrm{a}$ 
rapidly drops to zero when the distance becomes smaller than
20 nm,\cite{wokaun83,kuehn06a,anger06,ruppin82} whereas for two nanoparticles
it slightly increases and then decreases until
quenching (not shown), but
at shorter distances than for the previous case.
Thus, the data from Fig.~\ref{optical-antennas:antenna-design}a and b highlight
the competition between the SE rate enhancement and $\eta_\mathrm{a}$.
Note that the balance between them is clearly different for one
and two nanoparticles. 

\subsubsection{Changing the nanoparticle apex.~~}

To further discuss how the nanoparticle shape affects
the antenna performances,
we focus on the wavelength range between 600 nm and 
1100 nm, which covers the emission spectrum of relevant 
nanoscale light emitters.\cite{beveratos01,oconnell02,pavesi10,klimov10}
Because the enhancement is maximum when the emitter is placed at and oriented 
along the nanoparticle long axis, we only consider this situation.
Furthermore, to treat a more
experimentally feasible situation, we set the distance between emitter and
nanoparticle to 10 nm.

\begin{figure*}[!htb]
\centering{
\includegraphics[width=15cm]{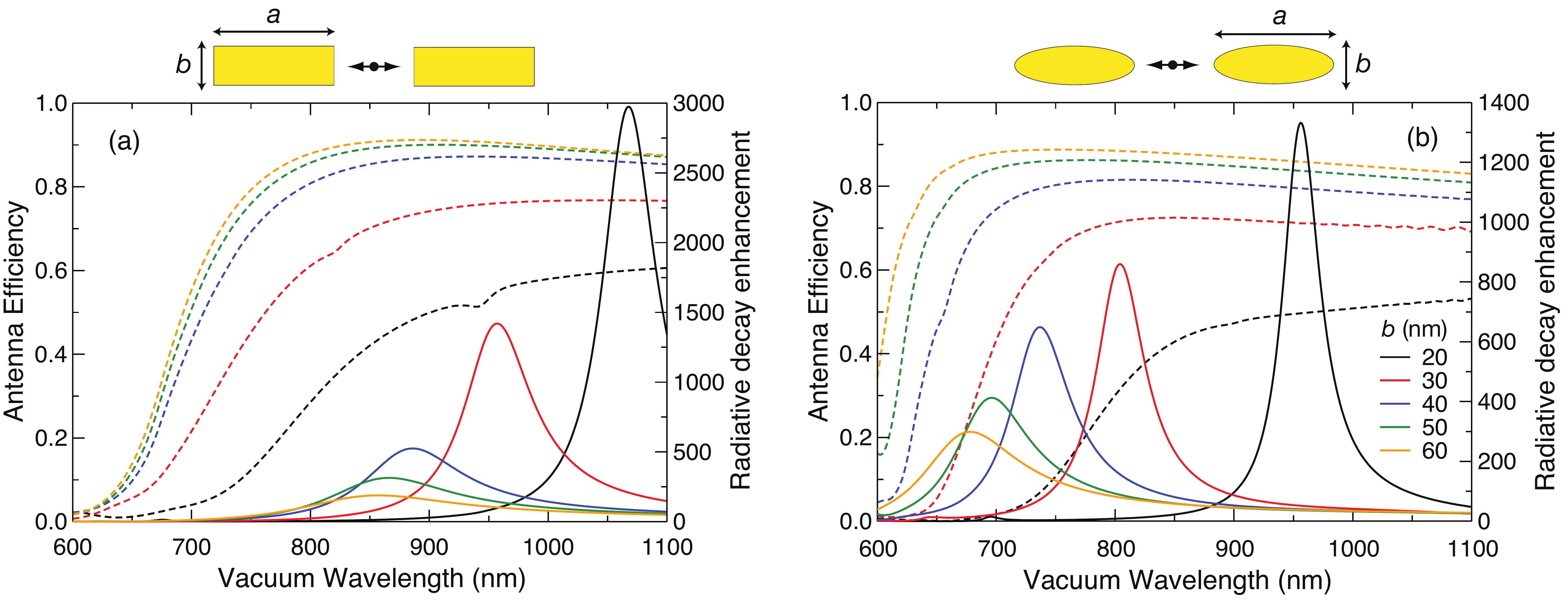}}
\caption{\label{optical-antennas:rods-spheroids}
Antenna efficiency (dashed curves) and radiative decay enhancement
(solid curves) for an emitter coupled to two gold nanorods (a) and
nanospheroids (b) in glass as a function of the
aspect ratio for $a$=80 nm. 
Figure adapted with permission from Ref.~\cite{mohammadi08b}.
Copyright (2008) by IOP Publishing.}
\end{figure*}

We compare
nanospheroids\cite{gersten81,klimov02} 
with gold nanorods.\cite{muehlschlegel05,aizpurua05}
Figure~\ref{optical-antennas:rods-spheroids}a shows that the
SPP resonance peak red shifts when the nanorod short axis $b$ decreases.
Therefore, by tuning the aspect ratio $a/b$ one can easily 
place the SPP resonance at the desired spectral 
location.\cite{aizpurua05} The increase of the radiative
decay rate is close to 3000 for wavelengths around 1100 nm.
However, when the resonance moves towards the visible spectrum,
the enhancement drops very rapidly.
As shown in Fig.~\ref{optical-antennas:rods-spheroids}a, $\eta_\mathrm{a}$
increases with the nanorod volume, i.e. as $b$ 
becomes larger. Unfortunately, the largest improvement correspond to 
the lowest efficiency because a higher aspect ratio implies a 
reduced volume.\cite{mohammadi09b} 

The steep decrease of the enhancement upon reduction of the aspect ratio
stems from the fact that the nanorods ends are flat.
In replacing nanorods with nanospheroids we identify three
important aspects. First, for high aspect ratios the nanospheroids
exhibit smaller enhancements. Second, for low aspect ratios the
enhancement decreases more slowly and the SPP resonance
is less red-shifted, as shown in Fig.~\ref{optical-antennas:rods-spheroids}b. 
Third, $\eta_\mathrm{a}$ reaches its plateau already 
at wavelengths close to 650 nm if the aspect ratio is less than 2.
Compared to nanorods the enhancement is larger at shorter wavelengths
because a smaller aspect ratio is partially compensated by a sharper
nanoparticle apex.
A more detailed comparison of the two antenna systems can be
found in Ref.~\cite{mohammadi08b}. These results highlight the fact that experiments
require great control over the 
nanoparticle shape, especially if large enhancements are desired.

\subsubsection{The conical antenna.~~}

An important issue is that $\eta_\mathrm{a}$
and the enhancement of SE are
maximal for different antenna parameters.
Can we improve the antenna design
to increase $\Gamma_\mathrm{t}$ without decreasing $\eta_\mathrm{a}$ and 
losing control on the spectral position of the resonance?
A simple solution is to use a nanocone, where one end
can be sharp to increase $K$ and the SE rate, whereas
the other end can be larger for increasing the volume,
hence $\eta_\mathrm{a}$.\cite{mohammadi10}

\begin{figure*}[!htb]
\centering{
\includegraphics[width=14cm]{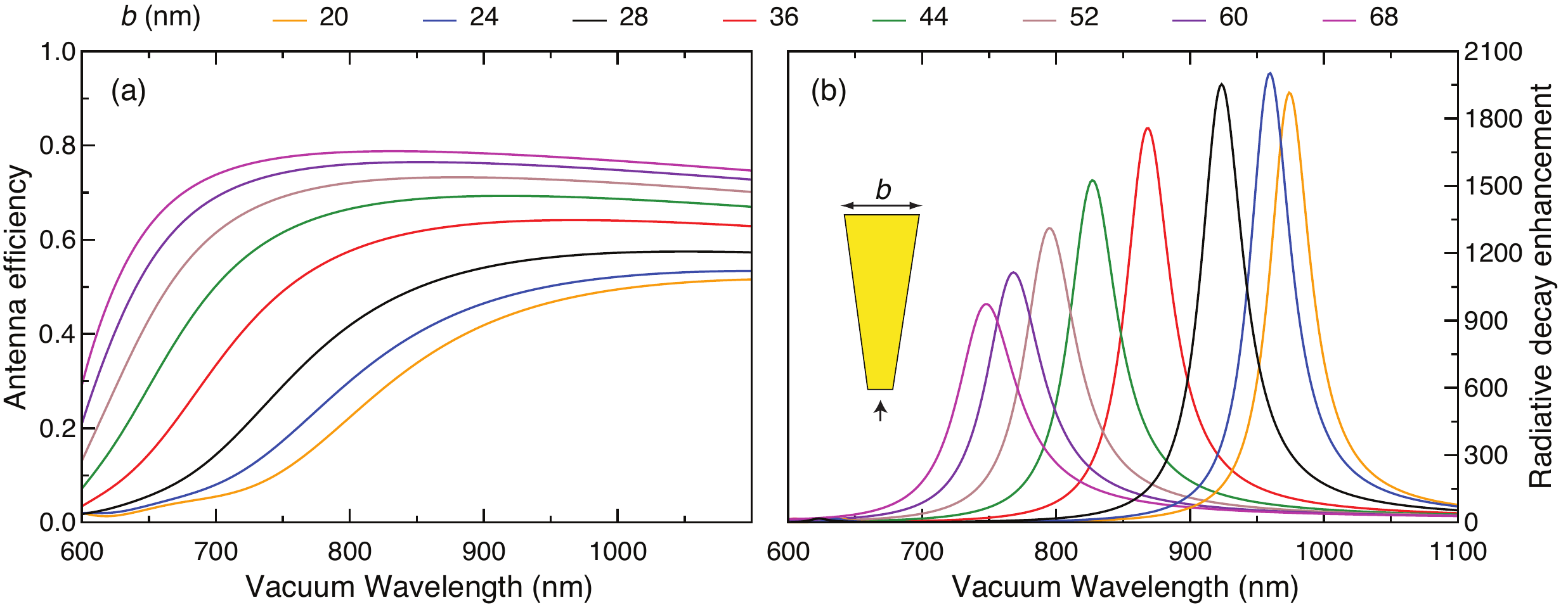}}
\caption{\label{optical-antennas:antenna-nanocone}
(a) Antenna efficiency and (b) radiative decay enhancement
as a function of the base diameter $b$.
The nanocone is in air, it is 140 nm long and it has a tip diameter of 20 nm.
Figure adapted with permission from Ref.~\cite{mohammadi10}.
Copyright (2010) by the American Chemical Society.}
\end{figure*}

Figure~\ref{optical-antennas:antenna-nanocone} displays the radiative decay
enhancement and $\eta_\mathrm{a}$ for single nanocones
as a function of the base diameter $b$.
The rate increases slightly and then
decreases, confirming that there exists an optimal value for
$b$.\cite{goncharenko06a,goncharenko07}
On the other hand, $\eta_\mathrm{a}$ grows with $b$ because the antenna volume
increases.
An important advantage with respect to nanorods and nanospheroids is
that here the resonance can be spectrally tuned 
by changing the nanocone angle, without a significant
loss of enhancement. 
Note that the enhancement factor is as high as 2000 for a conical
and 8000 for a bi-conical antenna (not shown).\cite{mohammadi10}

\subsection{Materials dependence}
\label{optical-antennas:material}

We have discussed examples where
the antenna properties are tuned by changing its shape and size.
While these degrees of freedom offer a wide range of performances,
there are situations where other parameters may be adjusted.
For instance, recent works have investigated 
the optical response of copper,\cite{tilaki07}
aluminum\cite{ekinci08b,langhammer08,chowdhury09,chan08,mohammadi09a}
and palladium\cite{pakizeh09b} nanoparticles. 
While previous theoretical studies focused on
$K$,\cite{cline86,zeman87} here
we discuss the modification of the SE rate and $\eta_\mathrm{a}$.
We choose nanospheroids as a model system 
and review designs that cover the spectral range
from the ultraviolet to the near-infrared.\cite{mohammadi09a,mohammadi09b}

\subsubsection{Background medium.~~}

\begin{figure}[h!]
\centering{
\includegraphics[width=7.5cm]{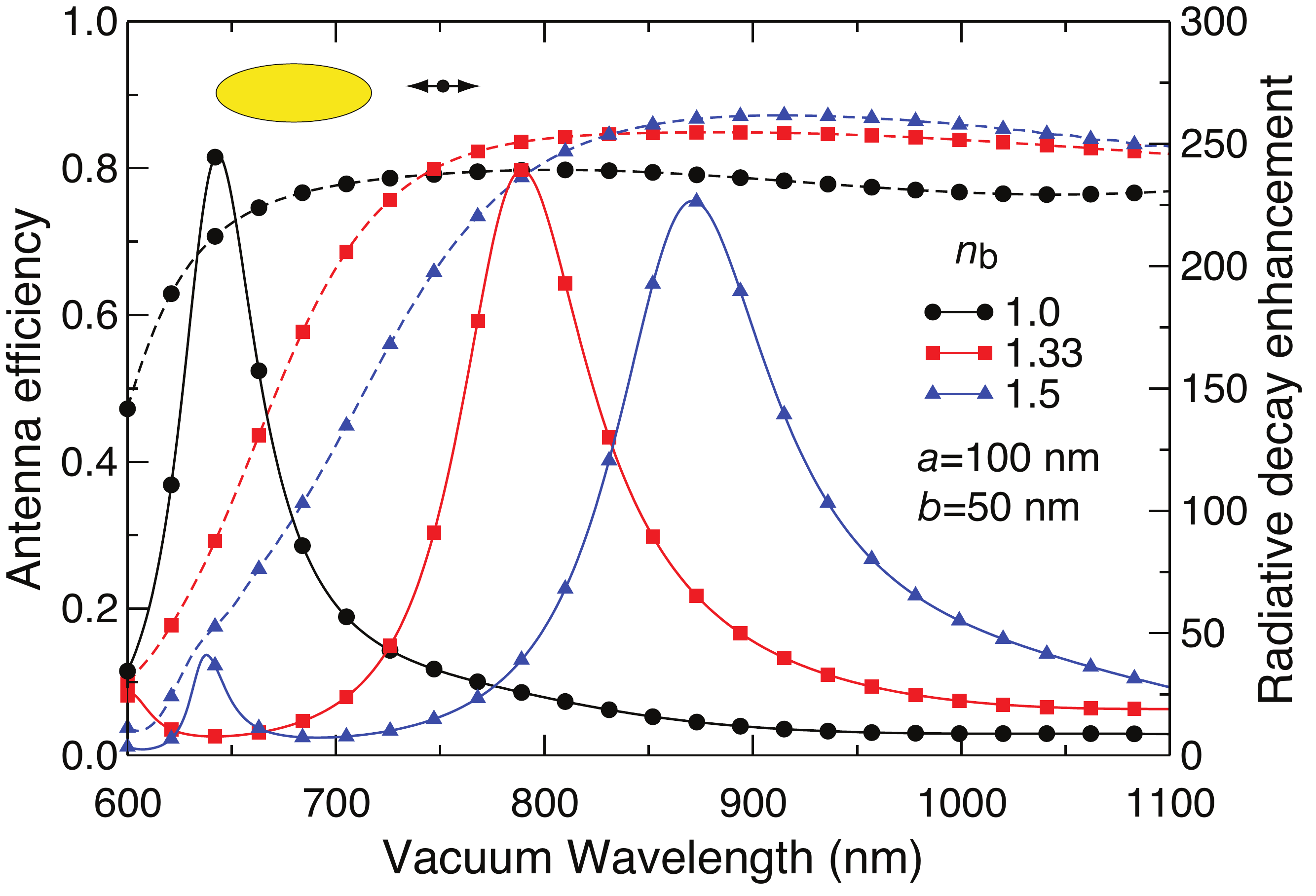}}
\caption{\label{optical-antennas:antenna-background}
Radiative decay enhancement (solid curves) and antenna
efficiency (dashed curves) for an emitter coupled to a gold nanospheroid.
Dependence on the background index $n_\mathrm{b}$.
Figure adapted with permission from Ref.~\cite{mohammadi09a}.
Copyright (2009) by the American Scientific Publishers.}
\end{figure}

First, we wish to illustrate how
the enhancement of the radiative decay rate and 
$\eta_\mathrm{a}$ depend on the background index.
Figure~\ref{optical-antennas:antenna-background}a shows that
for an emitter coupled to a gold nanospheroid
even a small change in the refractive index shifts
the SPP resonance by more than hundred nanometers.
At the same time, the resonance gets wider because
radiative broadening increases with the refractive index.\cite{wokaun82}
That also explains the small decrease in the enhancement.
Note that the shift of the SPP resonance towards shorter
wavelengths improves $\eta_\mathrm{a}$. For instance, it is 
larger than 70\% around 650 nm if the antenna is in air.

\subsubsection{Gold and copper.~~}

\begin{figure}[!htb]
\centering{
\includegraphics[width=7.5cm]{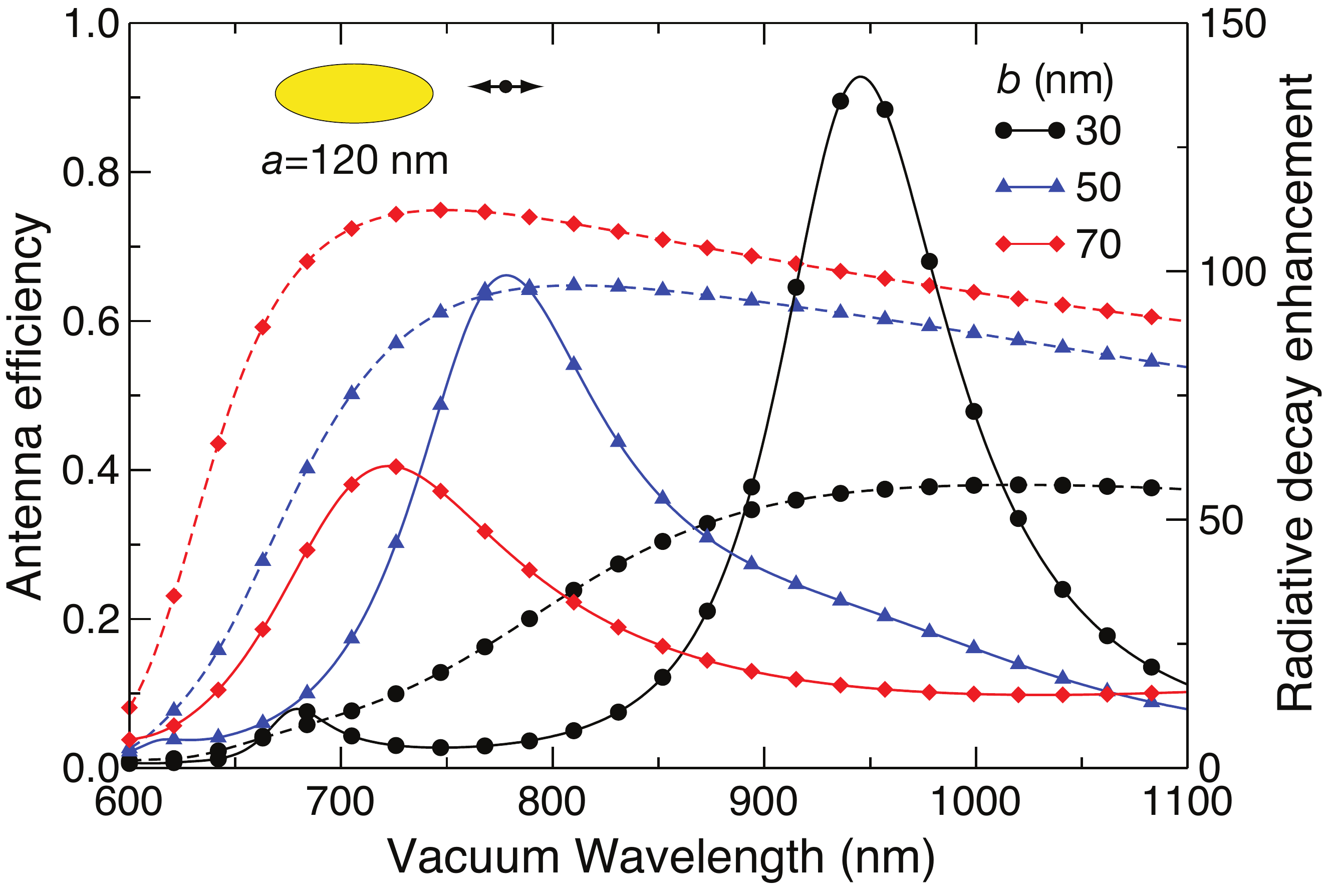}}
\caption{\label{optical-antennas:antenna-copper}
Radiative decay enhancement (solid curves) and antenna
efficiency (dashed curves) for an emitter coupled to
a copper nanospheroid in glass. 
Figure adapted with permission from Ref.~\cite{mohammadi09a}.
Copyright (2009) by the American Scientific Publishers.}
\end{figure}

The real part of the dielectric function of the two 
materials is quite similar, whereas the imaginary part is 
larger for copper.\cite{johnson72,CRChandbook} 
Figure~\ref{optical-antennas:antenna-copper} shows the radiative decay 
enhancement and $\eta_\mathrm{a}$ for an emitter 
coupled to a copper nanospheroid. Compared to gold 
antennas the enhancement is smaller and the resonances are broader, as 
expected from the larger imaginary part.
Moreover, $\eta_\mathrm{a}$ is lower, but it shows the same trend as
gold antennas. 

\subsubsection{Silver and aluminum.~~}

Silver has a higher plasma frequency than gold
so that the antenna resonance is shifted towards shorter wavelengths. 
On the other hand the imaginary part of the dielectric function drops
to lower values, with immediate benefits for
$\eta_\mathrm{a}$.\cite{johnson72}

Aluminum has an even higher plasma frequency than silver.\cite{palik98} 
Even if the imaginary part
is significantly larger than in the noble metals, in the region below 600 nm
the large and negative real part ensures that the skin depth
is sufficiently small to prevent significant absorption losses.

\begin{figure}[h!]
\centering{
\includegraphics[width=7.5cm]{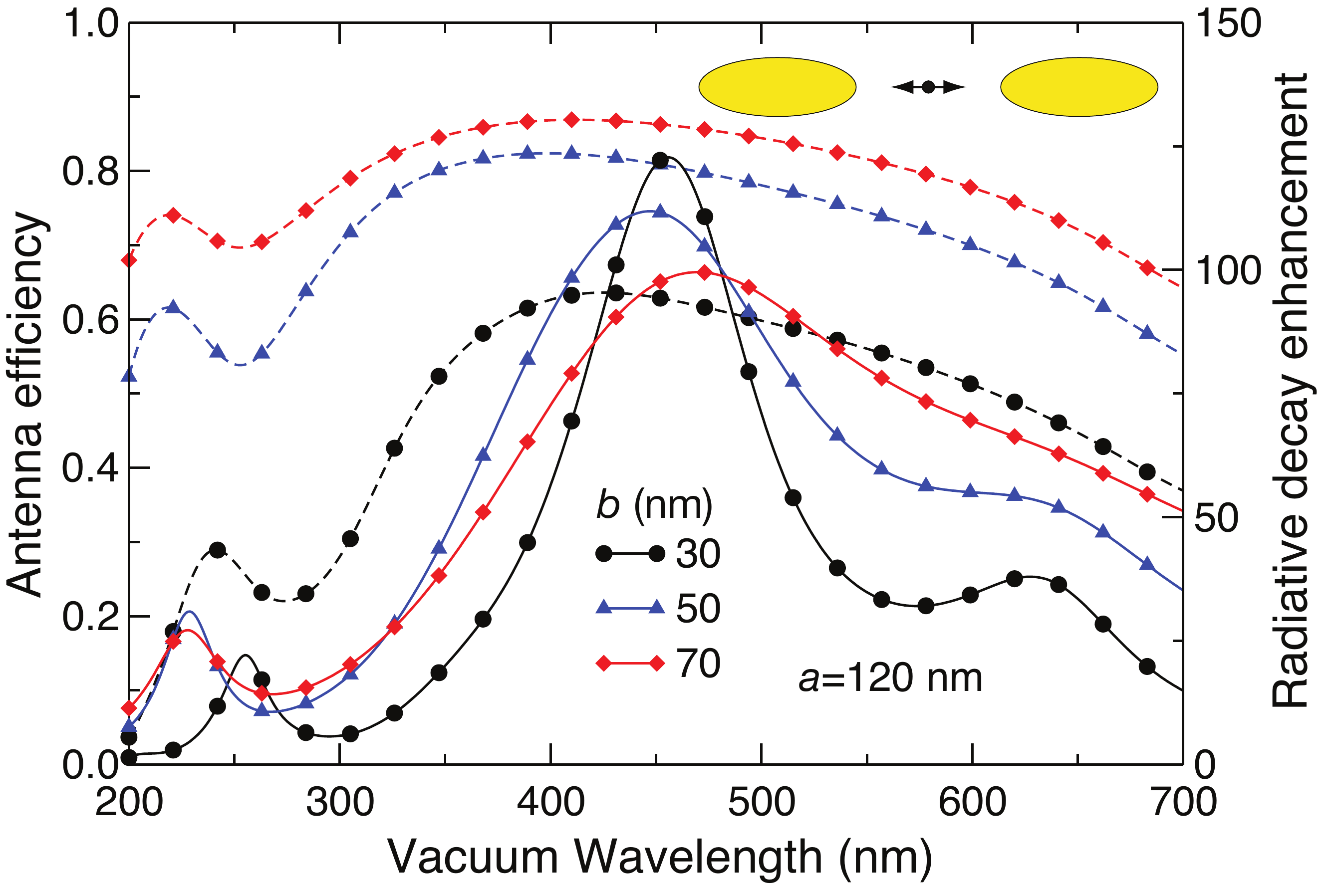}}
\caption{\label{optical-antennas:antenna-aluminum}
Radiative decay enhancement (solid curves) and antenna
efficiency (dashed curves) for an emitter coupled to
two aluminum nanospheroids in air. 
Figure adapted with permission from Ref.~\cite{mohammadi09a}.
Copyright (2009) by the American Scientific Publishers.}
\end{figure}

The antenna efficiency and the radiative decay enhancement for an 
emitter coupled to two aluminum nanospheroids
is provided in Fig.~\ref{optical-antennas:antenna-aluminum}.
The performances are not as high as for the same geometry 
made from other materials.\cite{mohammadi09a}
Since $\eta_\mathrm{a}$ is large the reason for that should be
attributed to radiative broadening rather than to losses.\cite{wokaun82}
Indeed, optimizations of $K$ have shown that the SPP 
resonance should be tuned around 200-300 nm.\cite{zeman87}
Aluminum is thus more suitable for applications
in the ultraviolet spectral region.\cite{krishanu07,chowdhury09}

\section{Towards nanoscale cavities}
\label{nanoscale-resonators:towards}

We have shown how optical antennas may
increase the SE rate by orders of magnitude
with minimal suffering from absorption losses.
These settings are very promising for implementing the
functionalities of microresonators at the nanoscale.
It is therefore instructive to translate the antenna performances
into the common parameters of an optical cavity, i.e. quality factor,
mode volume, Purcell factor and device footprint.

Recent works have discussed the
enhancement of light-matter interaction by optical antennas and
metal nanocavities using the mode-volume picture and the Purcell
factor.\cite{maier06b,oulton08,koenderink10,kuttge10}
Here we combine field-enhanced spectroscopy, antenna theory and cavity
QED to express figures of merit and scaling laws that
may provide useful insight on the opportunities offered
by optical antennas seen as nanoscale resonators.
Furthermore, we pay attention to the antenna efficiency and
study how it constrains the other performances.

\subsection{From antenna theory to nanoscale resonators}

First, we briefly review the formulation of cavity
QED in the perturbative regime, which is the
same level of theory used in the previous sections for
optical antennas. For convenience we set $\Gamma_\mathrm{nr}^o=0$,
and write $\Gamma_\mathrm{r}^o=\Gamma_o$ and $\Gamma_\mathrm{t}=\Gamma$.
Next, we discuss the relationship between the
Purcell factor and the modification of the SE
rate by optical antennas, with emphasis on the local density of
photonic states. We then establish a connection between
$K$ and the near-field zone of a radio-wave antenna.

\subsubsection{Cavity quantum electrodynamics.~~}

In free space the quantized electromagnetic field is expressed by
\begin{equation}
\label{nanoscale-resonators:QED}
\mathbf{E}(\mathbf{r})=i\sum_\mu\sqrt{\dfrac{\hbar\omega_\mu}{2\epsilon_0 V}}
\mathbf{e}_\mu\left(\hat{a}_\mu e^{i\mathbf{k\cdot r}}-\mathrm{h.c.}\right),
\end{equation}
where h.c. means Hermitian conjugation of the preceding term.
$V$ is the quantization volume, $\mathbf{e}_\mu$ is the polarization
versor and $\hat{a}_\mu$ is the destruction operator for one
photon in the mode $\mu$ of energy $\hbar\omega_\mu$.\cite{meystre07}
Using Eq.~(\ref{nanoscale-resonators:QED}) in Fermi golden
rule we obtain
\begin{equation}
\Gamma_o=\dfrac{2\pi}{\hbar}\sum_\mu
\dfrac{\hbar\omega_\mu}{2\epsilon_0V}|\mathbf{d}\cdot\mathbf{e}_\mu|^2
\delta(\hbar\omega-\hbar\omega_\mu)=
\dfrac{2\pi}{3\epsilon_0}\omega d^2g_o(\omega).
\end{equation}
$g_o(\omega)$ is the density of photonic states (DOS) in vacuo
for one polarization and it is given by
$g_o(\omega)=\omega^2/(2\pi^2\hbar c^3)$.

When the emitter is inside a resonator at $\mathbf{r}_o$,
Eq.~(\ref{nanoscale-resonators:QED})
needs to be replaced by\cite{meystre07}
\begin{equation}
\label{nanoscale-resonators:CQED}
\mathbf{E(r)}=i\sum_\mu\sqrt{\dfrac{\hbar\omega_\mu}{2\epsilon_0}}
\left(\hat{a}_\mu\bm{\alpha}_\mu(\mathbf{r})-\mathrm{h.c.}\right),
\end{equation}
where $\bm{\alpha}_\mu(\mathbf{r})$ is the cavity mode profile.
It is normalized to one and its dimensions correspond to $V^{-1/2}$.
Hence $|\bm{\alpha}_\mu(\mathbf{r})|^2$ may be viewed as
the probability density of having a photon at $\mathbf{r}$.
This intepretation will become apparent when we introduce the concept of
a mode volume to parametrize the enhancement of light-matter
interaction (see Eq.~(\ref{nanoscale-resonators:purcell})).
With Eq.~(\ref{nanoscale-resonators:CQED}) the SE rate becomes
\begin{equation}
\label{nanoscale-resonators:C-gamma}
\Gamma=\dfrac{2\pi}{\hbar}\sum_\mu\dfrac{\hbar\omega_\mu}{2\epsilon_0}
|\mathbf{d}\cdot\bm{\alpha}_\mu(\mathbf{r}_o)|^2\delta(\hbar\omega-\hbar\omega_\mu).
\end{equation}

We now assume that the transition frequency $\omega$ is resonant with
only one mode $\bm{\alpha}(\mathbf{r})$ and that $\mathbf{d}$
is parallel to the electric field. Next, the atomic line
is much narrower than the cavity mode and the latter has a Lorentzian
profile of width $\gamma$. Under these circumstances the DOS reads
\begin{equation}
g(\omega)=\dfrac{2}{\pi\hbar\gamma}=
\dfrac{2Q}{\pi\hbar\omega},
\end{equation}
where $Q=\omega/\gamma$ is the quality ($Q$) factor.
The mode volume for the position $\mathbf{r}_o$ is defined as
$V_\mu=|\bm{\alpha}(\mathbf{r}_o)|^{-2}$ and
Eq.~(\ref{nanoscale-resonators:C-gamma}) can be expressed in the form
\begin{equation}
\Gamma=\dfrac{2d^2Q}{\epsilon_0\hbar V_\mu}=F\Gamma_o,
\end{equation}
where F is the Purcell factor
\begin{equation}
\label{nanoscale-resonators:purcell}
F=\dfrac{3}{4\pi^2}\lambda^3\dfrac{Q}{V_\mu}.
\end{equation}
The condition for having a strong enhancement of the
SE rate is thus a high Q factor and a small $V_\mu$.
In place of $F$ one defines the local DOS (LDOS)
$\rho(\mathbf{r}_o,\omega)=g(\omega)|\bm{\alpha}(\mathbf{r}_o)|^2$
to express the SE rate as
\begin{equation}
\Gamma=\dfrac{\pi d^2\omega}{\epsilon_0}\rho(\mathbf{r}_o,\omega)=
\dfrac{\rho(\mathbf{r}_o,\omega)}{\rho_o(\mathbf{r}_o,\omega)}
\Gamma_o,
\end{equation}
where $\rho_o(\mathbf{r}_o,\omega)$ is the LDOS in vacuo.
Note that $V_\mu$ is often expressed in units of the cubic wavelength.
We do so in the following sections and write
$V_\mathrm{m}=(\lambda/n_\mathrm{b})^3V_\mu$, where the refractive index
$n_\mathrm{b}$ is added to generalize the formula to dielectric media.

\subsubsection{Field-enhanced spectroscopy.~~}

The theoretical models used for field-enhanced spectroscopy 
are based on the semi-classical theory of light-matter
interaction.\cite{metiu84,moskovits85}
Moreover, optical resonators are replaced by interfaces and
metal nanoparticles, which cannot be
easily described with the standard toolbox of
cavity QED.\cite{wylie86,knoll01,truegler08,savasta10}

The SE rate is thus computed from the expression
\begin{equation}
P_\mathrm{t}=-\dfrac{1}{2}\int_V\mathrm{Re}\left\{
\mathbf{j}^*(\mathbf{r},\omega)\cdot \mathbf{E}(\mathbf{r},\omega)
\right\}\mathrm{d}V,
\end{equation}
where $P_\mathrm{t}$ is the total power dissipated by the current
density $\mathbf{j}(\mathbf{r},\omega)$.\cite{jackson99}
For an infinitesimal oscillating
dipole $\mathbf{p}$ located at $\mathbf{r}_o$ one writes
$\mathbf{j}(\mathbf{r},\omega)=-i\omega\mathbf{p}
\delta(\mathbf{r-r}_o)$ and the previous equation takes the form
\begin{equation}
\label{nanoscale-resonators:Ptot-dipole}
P_\mathrm{t}=\dfrac{\omega}{2}\mathrm{Im}
\left\{\mathbf{p}^*\cdot\mathbf{E}(\mathbf{r}_o)\right\}.
\end{equation}

To make the connection with the modification of the LDOS
we recall that the electric
field radiated by $\mathbf{p}$ at $\mathbf{r}_o$ is related to the
Green tensor $\mathbf{G}$ by\cite{jackson99}
\begin{equation}
\mathbf{E(r)}=\dfrac{1}{\epsilon_0}\dfrac{\omega^2}{c^2}
\mathbf{G}(\mathbf{r},\mathbf{r}_o;\omega)\cdot\mathbf{p}
\end{equation}
and that\cite{chance78}
\begin{equation}
\label{nanoscale-resonators:LDOS}
\rho(\mathbf{r}_o,\omega)=\dfrac{6\omega}{\pi c^2}
\left[\mathbf{n}_\mathrm{p}\cdot\mathrm{Im}
\left\{\mathbf{G}(\mathbf{r}_o,\mathbf{r}_o;\omega)\right\}
\cdot\mathbf{n}_\mathrm{p}\right],
\end{equation}
where $\mathbf{n}_\mathrm{p}$ represents the dipole orientation.
By comparing Eqs.~(\ref{nanoscale-resonators:Ptot-dipole}) and
(\ref{nanoscale-resonators:LDOS}) we obtain
\begin{equation}
P_\mathrm{t}=\dfrac{\pi\omega^2}{12\epsilon_0}|\mathbf{p}|^2
\rho(\mathbf{r}_o,\omega),\hspace{0.5cm}\mathrm{and}\hspace{0.5cm}
\dfrac{P_\mathrm{t}}{P_o}=
\dfrac{\rho(\mathbf{r}_o,\omega)}{\rho_o(\mathbf{r}_o,\omega)}.
\end{equation}
Note that the change in the LDOS affects the total decay rate.
 
For an antenna that preserves the dipolar radiation pattern
of the emitter the modification of the radiative decay
rate can be related to $K$.\cite{taminiau08b}
To facilitate the derivation of analytical expressions
(see Sec.~\ref{nanoscale-resonators:figures-of-merit})
and gain insight on the various contributions
to $K$, we adopt a formalism based on polarizability models.
These have been extensively applied in the
1980s,\cite{metiu84,moskovits85} 
when it was difficult to perform electrodynamic analyses
on metal nanoparticles of arbitrary shape. 
For a free-space amplitude $E_o$, the
electric field near the antenna apex reads~\cite{wokaun82}
\begin{equation}
\label{nanoscale-resonators:Etip}
E_\mathrm{tip}=\xi E_\mathrm{dip}+E_o\simeq (1-L)\chi E_o,
\end{equation}
where $\xi$ represents the so-called lighting rod
effect and $E_\mathrm{dip}$ is the near field due to the electric
dipole induced in the antenna.\cite{moskovits85}
Equation~(\ref{nanoscale-resonators:Etip}) contains $\chi$,
the antenna susceptibility, and $L$, a geometrical
factor related to antenna shape.
Because $\Gamma_\mathrm{r}=\eta_\mathrm{a}\Gamma_\mathrm{t}$,
the change in the SE rate reads
\begin{equation}
\label{nanoscale-resonators:antenna-purcell}
\dfrac{\Gamma_\mathrm{t}}{\Gamma_o}\simeq
\left|\dfrac{E_\mathrm{tip}}{E_o}\right|^2\dfrac{1}{\eta_\mathrm{a}}.
\end{equation}

\subsubsection{Antenna theory.~~}
\label{nanoscale-resonators:antenna-theory}

Having established a relationship between the perturbative regime of 
cavity QED and the modification of the SE rate by optical antennas, we 
wish to investigate the connection between $K$ and antenna theory. We do 
so by considering the complex Poynting vector\cite{balanis05} 
\begin{equation} \mathbf{S}=\dfrac{1}{2}\mathbf{E\times H^*}, 
\end{equation} where $\mathbf{H}$ is the magnetic field. If we compute 
the power flow through a spherical surface of radius $r$, 
\begin{equation} \label{nanoscale-resonators:P-complex} 
P=\dfrac{1}{2}\int_{4\pi}\mathbf{S\cdot n}\,r^2\,\mathrm{d}\Omega = 
P_\mathrm{r}+iP_\mathrm{i}, \end{equation} we identify two terms. 
$P_\mathrm{r}$ is the power radiated by the antenna, whereas 
$iP_\mathrm{i}$ is purely imaginary and there is no time-average power 
flow associated with it. It is in fact called reactive power and it 
stands for the electromagnetic energy stored near the antenna. From 
Poynting theorem one can write \begin{equation} 
P_\mathrm{i}=2\omega(W_\mathrm{e}^\mathrm{r}-W_\mathrm{m}^\mathrm{r}), 
\end{equation} where $W_\mathrm{e}^\mathrm{r}$ and 
$W_\mathrm{m}^\mathrm{r}$ are the electric and magnetic energies in the 
radial direction, respectively.

The relationship between $P_\mathrm{i}$ and $K$
near an optical antenna can be understood by considering
the two quantities in Eq.~(\ref{nanoscale-resonators:P-complex})
for an infinitesimal dipole antenna, which read\cite{balanis05}
\begin{equation}
P_\mathrm{r}=Z\dfrac{\pi}{3}\left|\dfrac{I_ol}{\lambda}\right|^2,
\hspace{1cm}
P_\mathrm{i}=Z\dfrac{\pi}{3}
\left|\dfrac{I_ol}{\lambda}\right|^2\dfrac{1}{(kr)^3}.
\end{equation}
$l\ll \lambda$ is the dipole length, $I_o$ is the driving current,
$k=2\pi/\lambda$ and $Z$ is the vacuum impedance.

Note that $P_\mathrm{i}$ decreases with $kr$ and vanishes
in the far field, whereas $P_\mathrm{r}$ is constant.
Therefore, the reactive part of the antenna radiation field can
be associated with the field enhancement exhibited
by metal nanoparticles. Since these have dimensions smaller than
the wavelength, it turns out that near the metal surface  
$P_\mathrm{i}\gg P_\mathrm{r}$. By reciprocity, we can
argue that the incoming radiation becomes
reactive in the proximity of the nanoparticle
and it gives rise to a sizeable concentration of electromagnetic energy.

\subsubsection{Fundamental limitations.~~}

We now discuss some features starting from
electrically small antennas. Their name stems 
from the fact that the characteristic dimensions are much smaller than 
the wavelength of the field they radiate. Since antennas are devices 
conceived to couple to free space waves, one expects limitations upon 
size reduction. 

The theory of electrically small antennas
has been developed by several authors. Here we go after
the works of~\citeauthor{chu48}, \citeauthor{hansen81} and \citeauthor{mclean96} and focus on
the relationship between the $Q$ factor and
the reactive energy as a function of the antenna
dimensions.\cite{chu48,hansen81,mclean96}

The $Q$ factor can also be formulated as
\begin{equation}
Q=2\omega\dfrac{\max\{W_\mathrm{e},W_\mathrm{m}\}}{P_\mathrm{r}},
\end{equation}
where $W_\mathrm{e}$ and $W_\mathrm{m}$ are the time-averaged
electric and magnetic energies associated with the non-propagating part
of the electromagnetic field generated by the antenna. 
For electrically small antennas it turns out that $W_\mathrm{e}$
is much larger than $W_\mathrm{m}$, as expected for an oscillating
electric dipole.\cite{jackson99}

\citeauthor{chu48} considered an antenna enclosed in a virtual sphere
of radius $kr$ and computed the minimum $Q$ factor that it could have.
The calculation can be conveniently carried out by a multipole expansion of the
electromagnetic field, where $W_\mathrm{e}$ refers
to the non-propagating power external to the sphere.
For a linearly polarized antenna the theoretical minimum is given
by\cite{mclean96}
\begin{equation}
\label{nanoscale-resonators:Q-limit}
Q\simeq\eta_\mathrm{a}
\left(\dfrac{1}{(kr)^3}+\dfrac{1}{kr}\right),
\end{equation}
where we have added $\eta_\mathrm{a}$ to facilitate
the comparison with optical antennas. 

The $Q$ factor goes to infinity when $kr$ tends
to zero, meaning that an antenna cannot be
made indefinitely small without compromising its radiation
and bandwidth performances.
Note that for an infinitesimal dipole antenna with length $l=2r$
much larger than its cross section $2a$ the $Q$ factor,
\begin{equation}
\label{nanoscale-resonators:Q-dipole}
Q\simeq\dfrac{6\log(r/a)-1}{(kr)^2\tan(kr)},
\end{equation}
is larger than that of Eq.~(\ref{nanoscale-resonators:Q-limit}).
The dipole antenna exhibits worse performances because
it does not fully exploit the volume of the
virtual sphere.\cite{hansen81}

When an electrically small antenna approaches dimensions
where $kr \ll 1$, the $Q$ factor gets very large
and the system behaves as a subwavelength resonator.
It is worth pointing out that in
a microcavity the electromagnetic energy is prevented
from escaping into free space by
high-reflectivity mirrors, while here it is stored
because the antenna becomes a very inefficient radiator.

Interestingly, the increase in the $Q$ factor corresponds to a decrease 
in the antenna volume, which is also associated with $K$, as discussed in 
Sec.~\ref{nanoscale-resonators:antenna-theory}.
We therefore anticipate that the limitations of 
electrically small antennas become advantageous for enhancing the 
radiation properties of nearby quantum emitters.

\subsection{Figures of merit for optical antennas}
\label{nanoscale-resonators:figures-of-merit}

In Sec.~\ref{optical-antennas:chapter} we have presented 
antenna designs that could significantly improve light-matter
interaction. In place of rigorous electrodynamic calculations,
it is useful to present an approximate but sufficiently general model
that can be used to gain insight on these concepts and
to make the connection with antenna theory and cavity QED
in the perturbative regime.

\subsubsection{The model.~~}

We consider a prolate nanospheroid with long $a$ and
short $b$ semi-axes, whose physical volume is given by
$V_\mathrm{ph}=4\pi ab^2/3$. The antenna is made of a Drude
metal with dielectric function
\begin{equation}
\epsilon(\omega) = \epsilon_\mathrm{b}-
\dfrac{\omega_\mathrm{p}^2}{\omega(\omega+i\gamma)},
\end{equation}
where it is convenient to choose $\epsilon_\mathrm{b}$ equal
to that of the surrounding medium. $\omega_\mathrm{p}$ and $\gamma$
are the plasma and damping frequencies,
respectively.\cite{ashcroft76}
The optical properties of the antenna can be worked out starting
from a polarizability model with radiative corrections,\cite{wokaun82}
\begin{equation} 
\label{nanoscale-resonators:polarizability}
\alpha\simeq \dfrac{2\pi ab^2}{3L}\dfrac{\omega_o}
{\omega_o-\omega-i\Gamma_\mathrm{a}/2},
\end{equation}
where $\omega_o=\omega_\mathrm{p}\sqrt{L/\epsilon_\mathrm{b}}$ and
$\Gamma_\mathrm{a}=\gamma + 2k^3 ab^2 \omega_o /9L$ are the antenna
resonance frequency and linewidth, respectively.
Note that $\Gamma_\mathrm{a}$ has two contributions. The first term
represents absorption and the second one radiation. 
In Eq.~(\ref{nanoscale-resonators:polarizability}) we have introduced
the geometrical factor $L$, which is related to the aspect ratio
AR=$a/b$.\cite{bohren83b}
For a sphere AR=1 and $L=1/3$, while
for a prolate spheroid $L$ tends to 0 when $AR \gg 1$.

\begin{figure}[h!]
\centering{
\includegraphics[width=3.5cm]{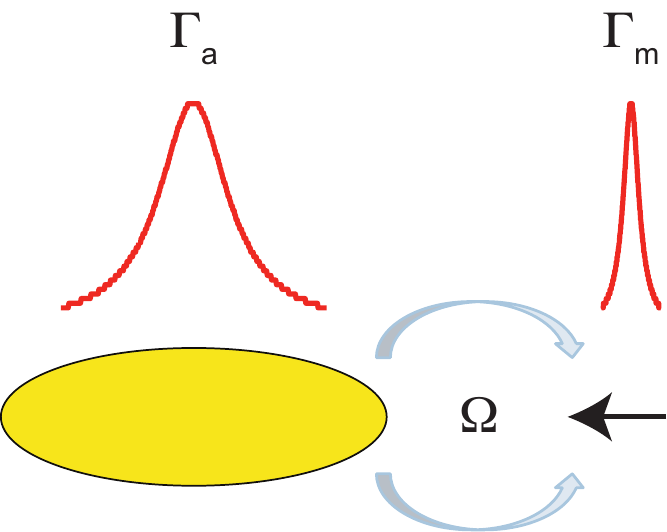}}
\caption{\label{nanoscale-resonators:model}
A quantum emitter (black arrow) coupled to an optical antenna
(gold spheroid) in the cavity QED picture.
$\Gamma_\mathrm{a}$ and $\Gamma_\mathrm{m}$ respectively represent
the antenna and the emitter resonance linewidths,
with $\Gamma_\mathrm{a}\gg \Gamma_\mathrm{m}$. The interaction
strength between the two systems is parametrized by the
Rabi frequency $\Omega$.}
\end{figure}

Figure~\ref{nanoscale-resonators:model} illustrates the antenna
model and the coupling to a quantum emitter. The latter has
the resonance frequency equal to that of the antenna, but the emitter linewidth
$\Gamma_\mathrm{m}$ is assumed to be much smaller than $\Gamma_\mathrm{a}$.
Furthermore, the interaction between 
the optical antenna and the emitter is formulated
using the vacuum Rabi frequency $\Omega$.\cite{meystre07}

\subsubsection{Antenna efficiency.~~}

We now derive the relevant antenna parameters starting from $\eta_\mathrm{a}$.
As discussed in Section~\ref{optical-antennas:chapter},
$\eta_\mathrm{a}$ depends on the antenna as well as on the position and
orientation of the emitter. To avoid details a good approximation
for $\eta_\mathrm{a}$ is the ratio between the scattering 
and the extinction
cross sections of the antenna. This definition
should not be considered a crude approximation, but rather an upper bound
that is very close to the realistic values obtained for high-performance
antennas.\cite{mohammadi09b}
Using Eq.~(\ref{nanoscale-resonators:polarizability}) we arrive at
\begin{equation}
\label{nanoscale-resonators:eta}
\eta_\mathrm{a}=\dfrac{1}{1+\dfrac{\gamma}{\omega_\mathrm{p}}
\dfrac{9\sqrt{\epsilon_\mathrm{b}L}\,\mathrm{AR}^2}{2(ka)^3}}.
\end{equation}
Equation~(\ref{nanoscale-resonators:eta}) shows
that $\eta_\mathrm{a}$ decreases quite rapidly
with the antenna volume, whereas the dependence on material losses enters
through the quantity $\gamma/\omega_\mathrm{p}$.\cite{ashcroft76}
Table~\ref{nanoscale-resonators:losses-metal} displays this
parameter for selected metals.
Note, however, that these values are for a static electric field.

\begin{table}[h!]
\small
\caption{\label{nanoscale-resonators:losses-metal}
Tabulated values of $\gamma/\omega_\mathrm{p}$ for selected
metals.\cite{ashcroft76} Lowering the temperature $T$ reduces the absorption losses
by an amount that is different for each metal.}
\begin{tabular*}{0.5\textwidth}{@{\extracolsep{\fill}}lll}
    \hline
    Material & $T=273K$ & $T=77K$ \\
    \hline
    Au & 0.0024 & 0.0006 \\
    Ag & 0.0018 & 0.00036 \\
    Al & 0.0051 & 0.00063 \\
    Cu & 0.0022 & 0.00029 \\
    \hline
  \end{tabular*}
\end{table}

\begin{figure}[h!]
\centering{
\includegraphics[width=6.5cm]{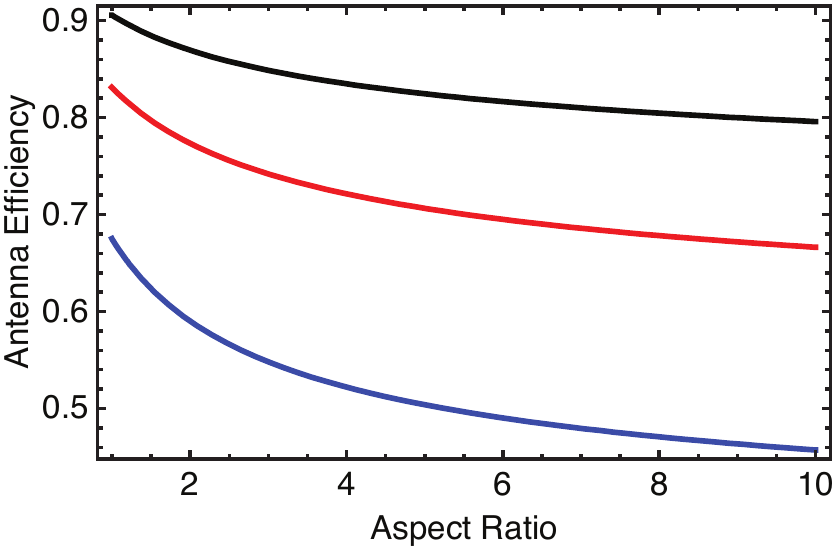}}
\caption{\label{nanoscale-resonators:efficiency-AR}
Antenna efficiency as a function of AR
plotted for three different values of $ka$:
0.5 (black), 0.4 (red), 0.3 (blue).
$\gamma/\omega_\mathrm{p}=0.005$ for all curves.}
\end{figure}

Figure~\ref{nanoscale-resonators:efficiency-AR} plots $\eta_\mathrm{a}$
as a function AR for different values of $ka$.
For a resonance wavelength of 600 nm, $ka=0.5$ corresponds to an optical
antenna with linear dimensions of the order of 100 nm.
Moreover, we choose $\gamma/\omega_\mathrm{p}=0.005$
and $\epsilon_\mathrm{b}=1$ to reproduce the performances
presented in Ref.~\cite{mohammadi08b}.
As expected, $\eta_\mathrm{a}$ decreases with AR and with
$ka$. Nonetheless, for $ka\simeq 0.5$ $\eta_\mathrm{a}$
is large in a wide range of aspect ratios.

\subsubsection{$Q$ factor.~~}

\begin{figure}[h!]
\centering{
\includegraphics[width=6.5cm]{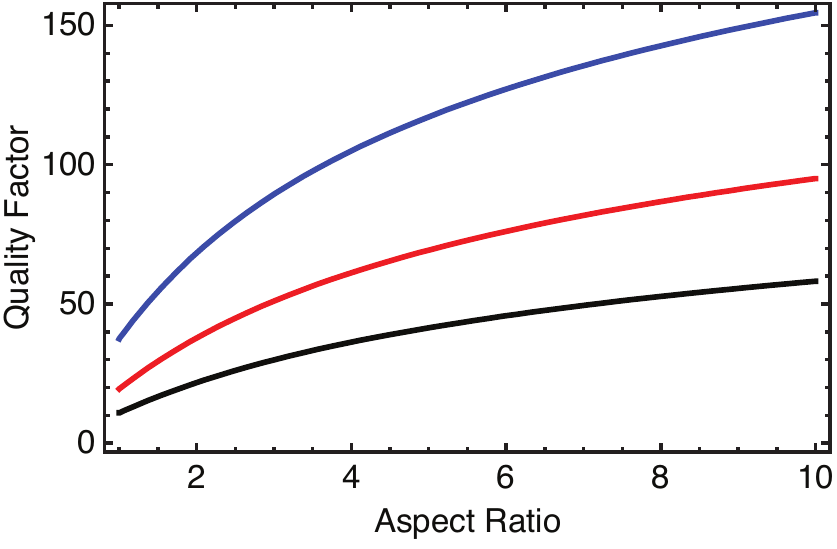}}
\caption{\label{nanoscale-resonators:quality-factor-AR}
$Q$ factor as a function of AR
plotted for three different values of $ka$:
0.5 (black), 0.4 (red), 0.3 (blue).
$\gamma/\omega_\mathrm{p}=0.005$ for all curves.}
\end{figure}

The $Q$ factor can be easily obtained from the formula
$Q=\omega_o/\Gamma_\mathrm{a}$.\cite{jackson99}
Adding $\eta_\mathrm{a}$ leads to
\begin{equation}
\label{nanoscale-resonators:Q-NP}
Q=\eta_\mathrm{a}\dfrac{9L\,\mathrm{AR}^2}{2(ka)^3}.
\end{equation}
Figure~\ref{nanoscale-resonators:quality-factor-AR}
displays the $Q$ factor as a function of AR
and $ka$. Note the competition between the decrease of
$\eta_\mathrm{a}$ in
Fig.~\ref{nanoscale-resonators:efficiency-AR} 
and the increase of the $Q$ factor with AR.
For $ka\ll 1$ absorption
losses dominate and the $Q$ factor saturates to the value 
$Q=(\omega_\mathrm{p}/\gamma)\sqrt{L/\epsilon_\mathrm{b}}$.

\subsubsection{Field enhancement.~~}

\begin{figure}[h!]
\centering{
\includegraphics[width=6.5cm]{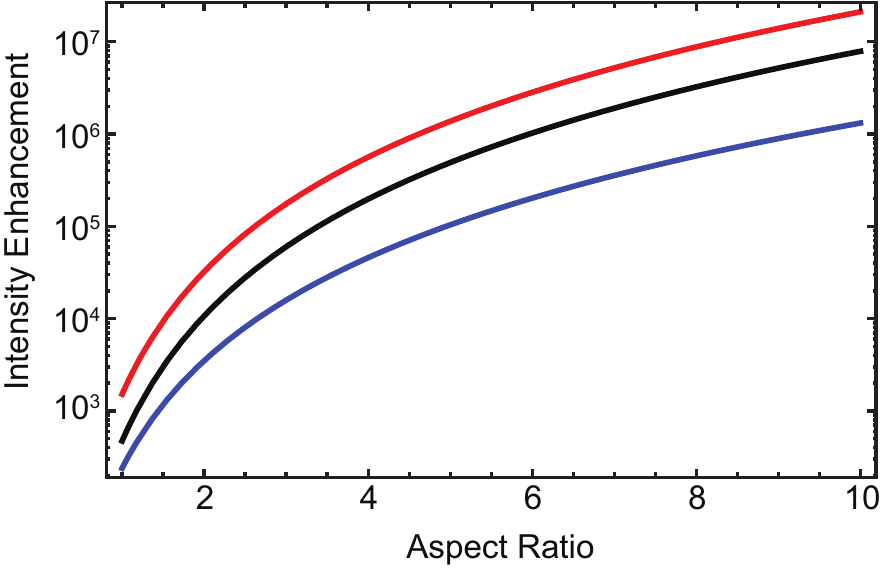}}
\caption{\label{nanoscale-resonators:field-enhancement-AR}
Intensity enhancement as a function of AR
plotted for different values of $ka$ and $\gamma/\omega_\mathrm{p}$:
$ka=0.5$ and $\gamma/\omega_\mathrm{p}=0.005$ (black),
$ka=0.4$ and $\gamma/\omega_\mathrm{p}=0.005$ (red),
$ka=0.4$ and $\gamma/\omega_\mathrm{p}=0.05$ (blue).}
\end{figure}

For the calculation of $K$ we consider the antenna apex.
We start from Eq.~(\ref{nanoscale-resonators:Etip})
and replace $\xi$ and $E_\mathrm{dip}$ with the values
obtained from Eq.~(\ref{nanoscale-resonators:polarizability}).
The lighting-rod effect reads
$\xi=3AR^2(1-L)/2$ and $E_\mathrm{dip}=2\alpha E_o/a^3$.
A few algebraic operations lead to
\begin{equation}
\label{nanoscale-resonators:K}
K=\left(\dfrac{9}{2}\eta_\mathrm{a}\dfrac{\mathrm{AR}^2}{(ka)^3}\right)^2
(1-L)^2.
\end{equation}
Note that the $(ka)^{-6}$ dependence is compensated by a drop in $\eta_\mathrm{a}$. 
In fact, when $ka$ approaches zero, $K$ saturates to the value
\begin{equation}
\lim_{ka\to 0}K=\left(\dfrac{\omega_\mathrm{p}}{\gamma}\right)^2
\dfrac{(1-L)^2}{\epsilon_\mathrm{b}L},
\end{equation}
which depends on the material losses and the antenna geometry.
Indeed, Fig.~\ref{nanoscale-resonators:field-enhancement-AR} indicates
that $K$ falls off when $\gamma/\omega_\mathrm{p}$ increases.

\subsubsection{Optical antennas are electrically small.~~}

\begin{figure}[h!]
\centering{
\includegraphics[width=6.5cm]{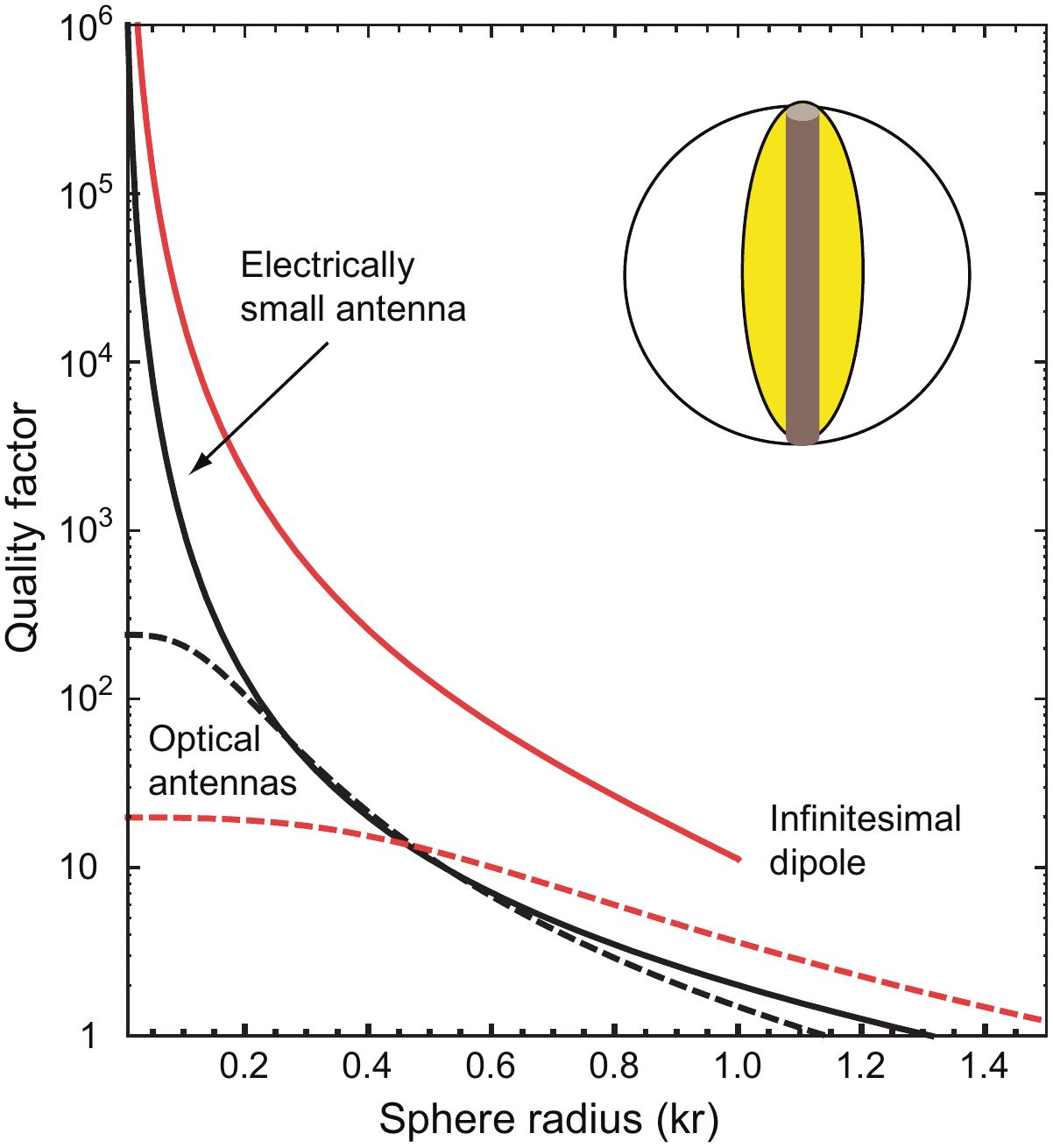}}
\caption{\label{nanoscale-resonators:antenna-q-factor}
$Q$ factor as a function of $kr$
for an ideal electrically small antenna
(black solid curve), an infinitesimal dipole (red solid curve),
and optical antennas (dashed curves): a
nanosphere (black) and a nanospheroid (red). 
We choose $\gamma/\omega_\mathrm{p}=0.003$ for the nanosphere,
AR=3 and $\gamma/\omega_\mathrm{p}=0.05$ for the nanospheroid,
and $r/a=50$ for the infinitesimal dipole.
The inset shows an infinitesimal dipole (brown)
and a nanospheroid (yellow) enclosed in the radiating sphere.}
\end{figure}

The inset in Fig.~\ref{nanoscale-resonators:antenna-q-factor}
depicts a nanospheroid and an infinitesimal dipole
antenna enclosed in the Chu virtual sphere of radius $r$.
We also consider a nanosphere and an ideal electrically small antenna.
The $Q$ factor for these radiating systems is plotted in
Fig.~\ref{nanoscale-resonators:antenna-q-factor}
as a function of $kr$.
According to the Chu theory,
a metal nanosphere should be an efficient electrically
small antenna, because it can fill the virtual sphere.
Indeed the $Q$ factor of a nanosphere agrees very well
with the result of Eq.~(\ref{nanoscale-resonators:Q-limit})
when $kr <1 $. However, for $kr \ll 1$ the curve saturates to
$\omega_\mathrm{p}/(\sqrt{3}\gamma)$.
When the nanosphere is replaced by a nanospheroid the
$Q$ factor increases, because the available radiating volume
is not fully exploited.

In summary, metal nanoparticles are electrically small antennas,
agree with the Chu theory and share the resulting limitations.
These turn out to be very important for optical antennas, because
the fact that the $Q$ factor and the reactive
energy increase when the antenna volume decreases
may be exploited to enhance light-matter interactions.

\subsection{Comparison with optical resonators}

\begin{figure*}[!htb]
\centering{
\includegraphics[width=12cm]{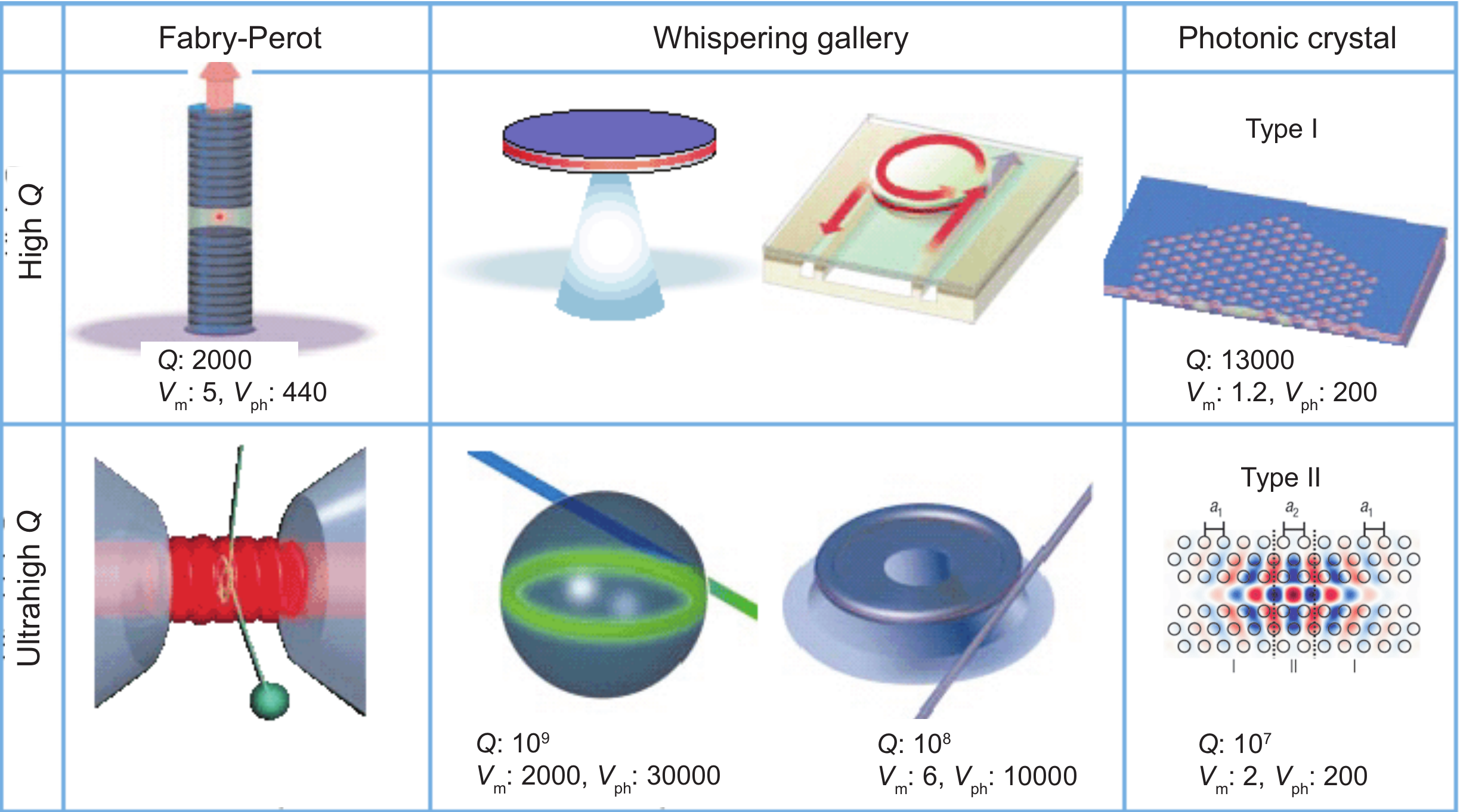}}
\caption{\label{nanoscale-resonators:resonators}
Figures of merit for optical resonators: $Q$ factor,
mode volume $V_\mathrm{m}$ and physical volume $V_\mathrm{ph}$.
$V_\mathrm{m}$ and $V_\mathrm{ph}$ are both in units of the cubic
wavelength.
Figure adapted with permission from Macmillan Publishers Ltd: 
Nature (Ref.~\cite{vahala03}), copyright (2003) and
Nat. Mater (Ref.~\cite{song05}), copyright (2005).}
\end{figure*}

We are ready to compare the figures of merit of optical antennas
with those of optical microcavities.
For our purpose we choose the following cavity
parameters: radiation efficiency, $Q$ factor, mode volume and footprint.
The latter represents the actual device volume $V_\mathrm{ph}$.
Literature values for these quantities are indicated
in Fig.~\ref{nanoscale-resonators:resonators} with
the corresponding resonator models.\cite{vahala03,song05}

\subsubsection{Antenna efficiency.~~}

For a more direct comparison with optical resonators, we use
$V_\mathrm{ph}$ in units of $(\lambda/n_\mathrm{b})^3$ to obtain
\begin{equation}
\eta_\mathrm{a}=\dfrac{1}{1+\dfrac{\gamma}{\omega_\mathrm{p}}
\dfrac{3}{4\pi^2}\dfrac{\sqrt{\epsilon_\mathrm{b}L}}{V_\mathrm{ph}}}.
\end{equation}

\begin{figure}[h!]
\centering{
\includegraphics[width=6.25cm]{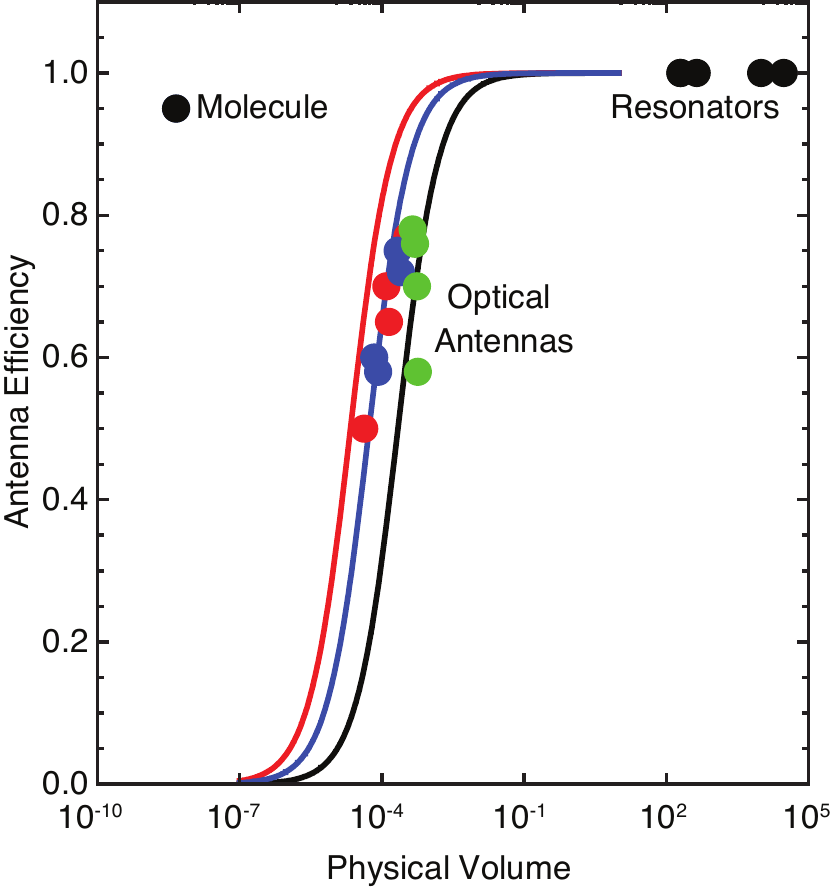}}
\caption{\label{nanoscale-resonators:antenna-efficiency}
Antenna efficiency as a function of $V_\mathrm{ph}$.
The curves represent the antenna model:
AR=1, $\gamma/\omega_\mathrm{p}=0.005$ (black),
AR=10, $\gamma/\omega_\mathrm{p}=0.005$ (blue),
AR=1, $\gamma/\omega_\mathrm{p}=0.0005$ (red).
The circles refer to resonators, a molecule and optical
antennas:\cite{mohammadi08b} nanospheroids (green),
nanorods (red), and nanorod pairs (blue).}
\end{figure}

The curves plotted in Fig.~\ref{nanoscale-resonators:antenna-efficiency}
correspond to different values of $\gamma/\omega_\mathrm{p}$ and AR
(see the figure caption for details).
It is shown that $\eta_\mathrm{a}$ drops when $V_\mathrm{ph}$ is smaller
than about 10$^{-4}$ cubic wavelengths, a value that strongly depends on
$\gamma/\omega_\mathrm{p}$.
On top of these curves the filled circles refer to antenna designs
discussed in Ref.~\cite{mohammadi08b}, namely
nanospheroids (green), nanorods (red) and nanorod pairs (blue).
The data agree well with our model. The dependence of $\eta_\mathrm{a}$
on $V_\mathrm{ph}$ illustrates the competition between absorption and
radiation losses and recalls the conflict with the enhancement of
light-matter interaction, which requires an optical antenna with
a strong reactive behavior. For the sake of comparison, we also
indicate $V_\mathrm{ph}$ and $\eta_\mathrm{a}$ for optical resonators
and a molecule with $\eta_o\simeq 1$.

\subsubsection{$Q$ factor.~~}

The $Q$ factor is inversely proportional to $V_\mathrm{ph}$
and Eq.~(\ref{nanoscale-resonators:Q-NP}) can be rewritten as
\begin{equation}
\label{nanoscale-resonators:Q-factor}
Q=\eta_\mathrm{a}\dfrac{3}{4\pi^2}\dfrac{L}{V_\mathrm{ph}}
\end{equation}

\begin{figure}[h!]
\centering{
\includegraphics[width=6.25cm]{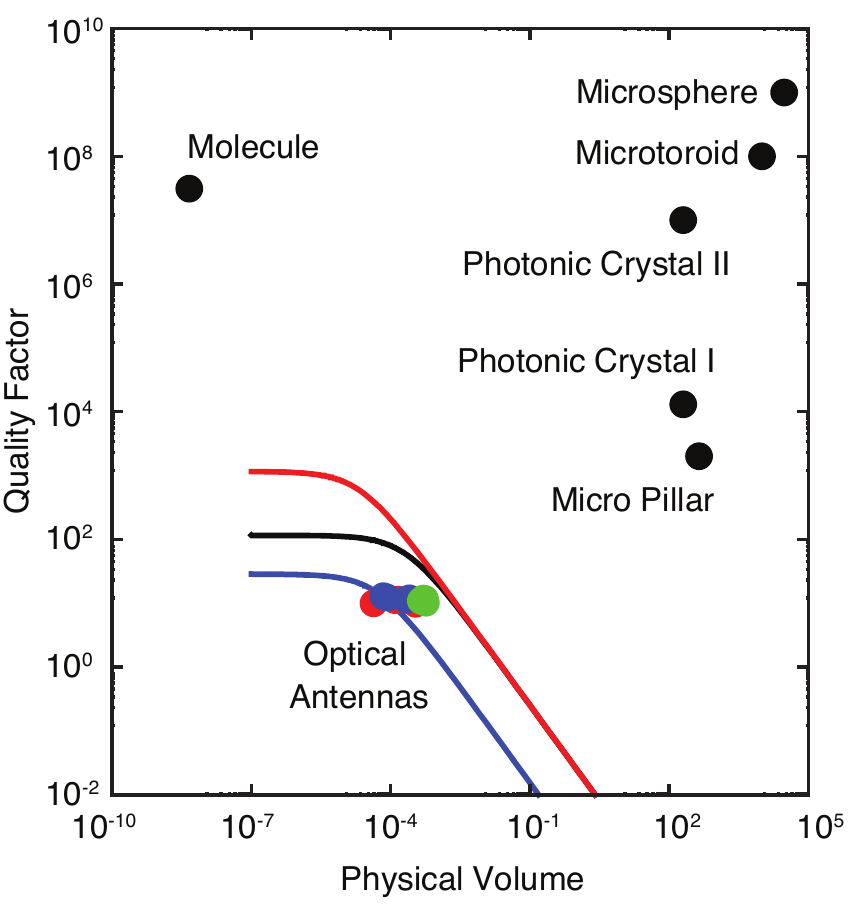}}
\caption{\label{nanoscale-resonators:quality-factor}
$Q$ factor as a function of $V_\mathrm{ph}$.
The curves represent the antenna model:
AR=1, $\gamma/\omega_\mathrm{p}=0.005$ (black),
AR=10, $\gamma/\omega_\mathrm{p}=0.005$ (blue),
AR=1, $\gamma/\omega_\mathrm{p}=0.0005$ (red).
The filled circles refer to resonators, a molecule
and optical antennas:\cite{mohammadi08b} nanospheroids (green),
nanorods (red), and nanorod pairs (blue).}
\end{figure}

Figure~\ref{nanoscale-resonators:quality-factor} compares
Eq.~(\ref{nanoscale-resonators:Q-factor}) with the antenna designs
of Ref.~\cite{mohammadi08b}, as well as with selected
optical resonators and a single molecule.
The $Q$ factor of optical antennas is much smaller than in the other
systems and for very small values of $V_\mathrm{ph}$ it is determined
by the absorption losses and the antenna geometry. Since the
response time is proportional to the $Q$ factor, optical antennas
might represent a unique opportunity for enhancing light-matter
interaction and, at the same time, meet the requirements of
ultrafast optics. For example, a single molecule or ultrahigh-$Q$
cavities have response times of the order of
nanoseconds. Resonators with a high $Q$ factor
can cope with picosecond pulses. Optical antennas
may offer the possibility of working with femtosecond pulses.
In this respect, an important point of concern is whether
antennas could increase light-matter interaction as much as
optical resonators.

\subsubsection{Spontaneous emission rate.~~}

The enhancement of the SE rate is obtained
from Eq.~(\ref{nanoscale-resonators:antenna-purcell}) upon
replacing $K$ with the expression given in
Eq.~(\ref{nanoscale-resonators:K}). A few more algebraic steps lead to
\begin{equation}
\label{nanoscale-resonators:spont-emission}
\dfrac{\Gamma_\mathrm{t}}{\Gamma_o}=
\eta_\mathrm{a}\dfrac{9}{16\pi^4}\dfrac{(1-L)^2}{V_\mathrm{ph}^2}.
\end{equation}

\begin{figure}[h!]
\centering{
\includegraphics[width=6.25cm]{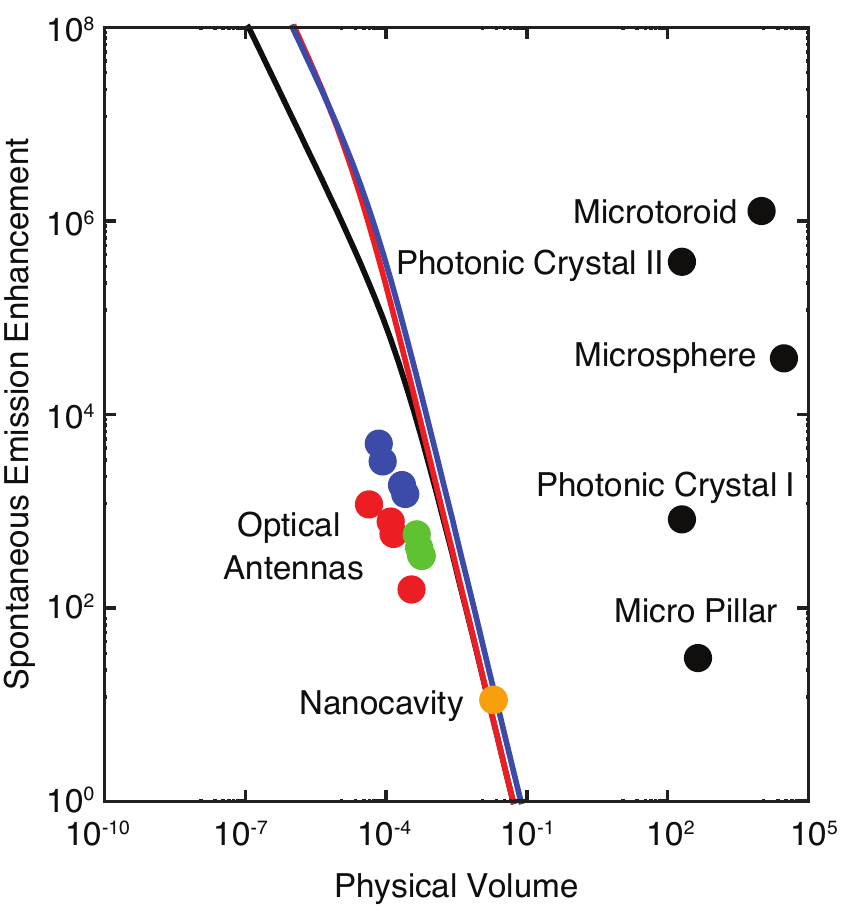}}
\caption{\label{nanoscale-resonators:spontaneous-emission}
Enhancement of the SE rate as a function of $V_\mathrm{ph}$.
The curves represent the antenna model:
AR=1, $\gamma/\omega_\mathrm{p}=0.005$ (black),
AR=10, $\gamma/\omega_\mathrm{p}=0.005$ (blue),
AR=1, $\gamma/\omega_\mathrm{p}=0.0005$ (red).
The filled circles refer to resonators, a nanocavity\cite{maksymov10}
and optical antennas:\cite{mohammadi08b} nanospheroids (green),
nanorods (red), and nanorod pairs (blue).}
\end{figure}

Figure~\ref{nanoscale-resonators:spontaneous-emission} demonstrates
that the Purcell factor of optical resonators and the modification
of the SE rate by optical antennas can be
of the same order of magnitude. Furthermore, it is shown that
the designs discussed in Ref.~\cite{mohammadi08b}
can compete with the performances of high-$Q$ photonic-crystal cavities

\subsubsection{Mode volume.~~}

The last topic to be discussed is the mode volume.
We point out that $V_\mathrm{m}$ is not a well defined quantity
for optical antennas, because a dissipative environment does
not have normalizable true modes.\cite{dutra00} 
Since we are mostly interested in presenting figures of merit
and scaling laws, we are satisfied with a
definition of $V_\mathrm{m}$ based on Eq.~(\ref{nanoscale-resonators:purcell}). 
We thus write
\begin{equation}
V_\mathrm{m}=\dfrac{3}{4\pi^2}Q\left(\dfrac{\Gamma_\mathrm{t}}{\Gamma_o}\right)^{-1}.
\end{equation}
We then replace the $Q$ factor and the enhancement of the SE
rate using Eqs.~(\ref{nanoscale-resonators:Q-factor}) and
(\ref{nanoscale-resonators:spont-emission}), respectively, to arrive at
\begin{equation}
\label{nanoscale-resonators:mode-antenna}
V_\mathrm{m}=\dfrac{L}{(1-L)^2}V_\mathrm{ph}.
\end{equation}

\begin{figure}[h!]
\centering{
\includegraphics[width=6.25cm]{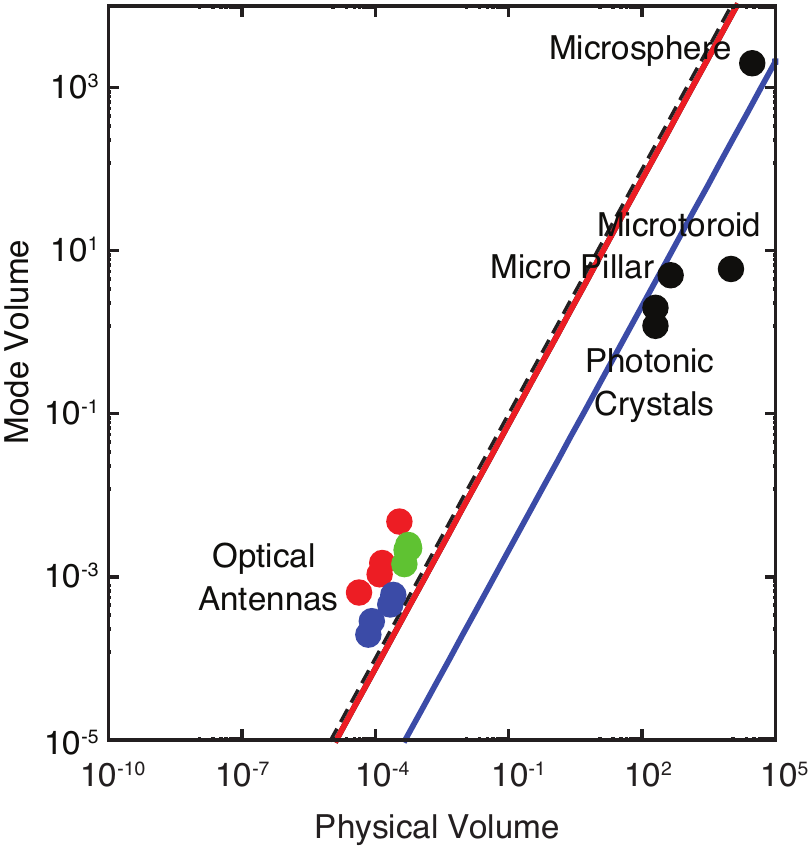}}
\caption{\label{nanoscale-resonators:mode-volume}
Mode volume as a function of $V_\mathrm{ph}$.
The curves represent the antenna model:
AR=1, $\gamma/\omega_\mathrm{p}=0.005$ (black),
AR=10, $\gamma/\omega_\mathrm{p}=0.005$ (blue),
AR=1, $\gamma/\omega_\mathrm{p}=0.0005$ (red).
The filled circles refer to resonators and
optical antennas:\cite{mohammadi08b}
nanospheroids (green), nanorods (red), and nanorod pairs (blue).
The dashed line marks $V_\mathrm{m}=V_\mathrm{ph}$.}
\end{figure}

Figure~\ref{nanoscale-resonators:mode-volume} compares the
result of Eq.~(\ref{nanoscale-resonators:mode-antenna})
with the mode volume of optical resonators.
Note that even for the smallest photonic-crystal
cavities $V_\mathrm{m}$ is about three orders of magnitude
larger than for optical antennas. 
Furthermore, while for the latter $V_\mathrm{m}$ is comparable to
$V_\mathrm{ph}$, for microcavities $V_\mathrm{ph}$ is significantly
larger than $V_\mathrm{m}$.

An alternative way to derive $V_\mathrm{m}$ for an optical antenna
utilizes the vacuum Rabi frequency $\Omega$. The latter can be obtained
from a Green-function formulation of QED.\cite{wylie86,knoll01}
If the antenna leads to a strong modification of the SE
rate ($\Gamma_\mathrm{t}/\Gamma_o \gg 1$),
we can ignore the free-space radiation modes
and approximate the imaginary part of the Green function with
a Lorentzian of width $\Gamma_\mathrm{a}$.
It can be shown that the Rabi frequency is related
to $\Gamma_\mathrm{t}$ and $\Gamma_\mathrm{a}$ through the formula
\begin{equation}
\Omega=\sqrt{\dfrac{\Gamma_\mathrm{t}\Gamma_\mathrm{a}}{4}}.
\end{equation}
From Eqs.~(\ref{nanoscale-resonators:polarizability}) and
(\ref{nanoscale-resonators:spont-emission}) we find
\begin{equation}
\Omega=\dfrac{1-L}{\sqrt{L}}\dfrac{1}{\sqrt{V_\mathrm{ph}}},
\end{equation}
Note that the above expression is given in units of
$\omega^2 d/(4\sqrt{\epsilon_0\hbar\pi^3 c^3})$.
Since $\Omega=\sqrt{\omega_o d^2/(2\epsilon_0\hbar V_\mu)}$, where
$V_\mu$ is the mode volume in dimensional units, one obtains
the same result of Eq.~(\ref{nanoscale-resonators:mode-antenna}).

\begin{figure}[h!]
\centering{
\includegraphics[width=5cm]{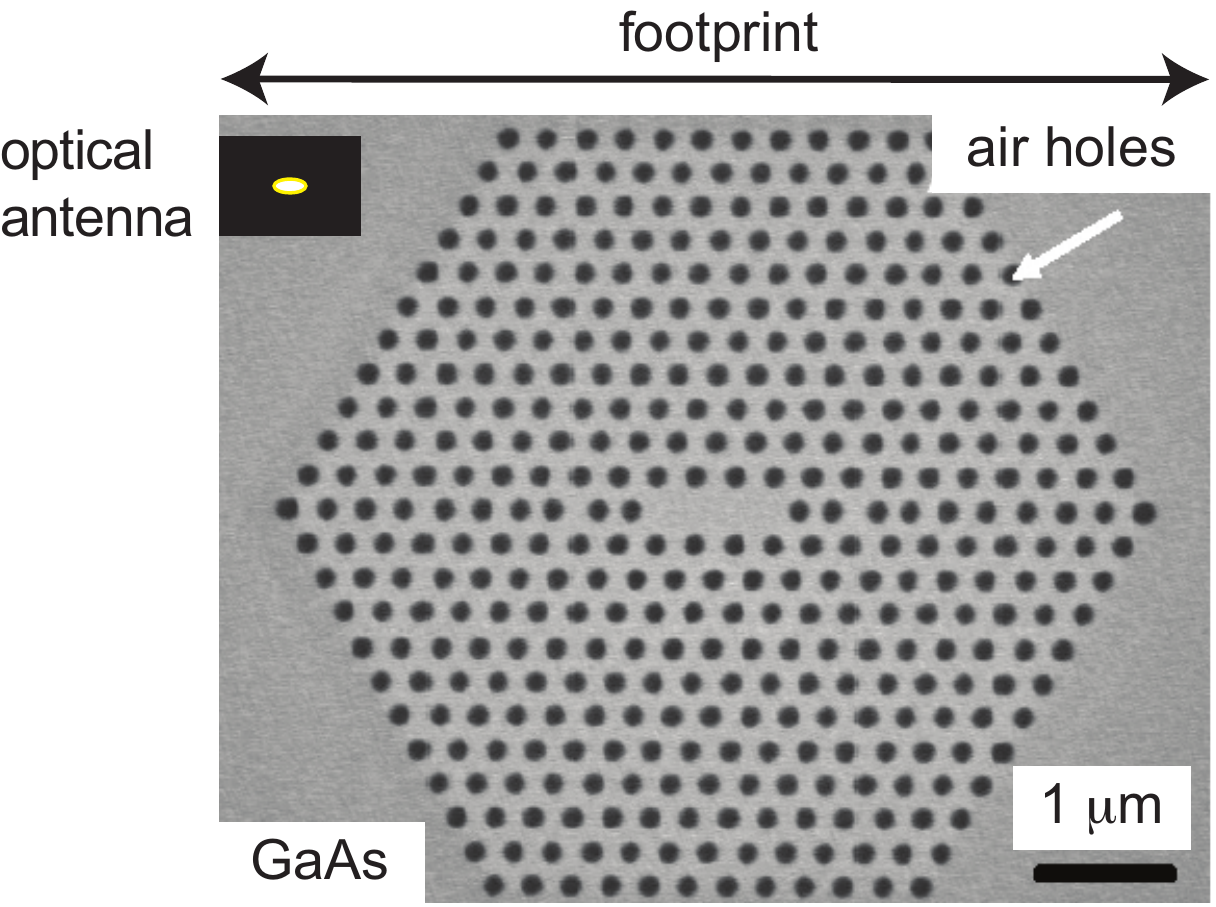}}
\caption{\label{nanoscale-resonators:footprint}
Device footprint for a GaAs
photonic-crystal cavity (figure adapted with permission from Ref.~\cite{nomura07}.
Copyright (2007) by SPIE) and an optical antenna.}
\end{figure}

After these considerations, we once more wish to discuss
the competition between $\eta_\mathrm{a}$ and the enhancement
of light-matter interaction.
While $\eta_\mathrm{a}$ drops very rapidly when the antenna dimensions
become smaller than a certain value that primarily depends on the
parameter $\gamma/\omega_\mathrm{p}$, the enhancement of the SE rate
increases and, despite the low $Q$ factor, it reaches
values that compete with those of state-of-the-art optical cavities.
Within these opposite trends there is a parameter range where
optical antennas could function as nanoscale resonators with
a tiny device footprint (see Fig.~\ref{nanoscale-resonators:footprint}),
manageable absorption losses and ultrafast operation.

\begin{figure*}[!htb]
\centering{
\includegraphics[width=12cm]{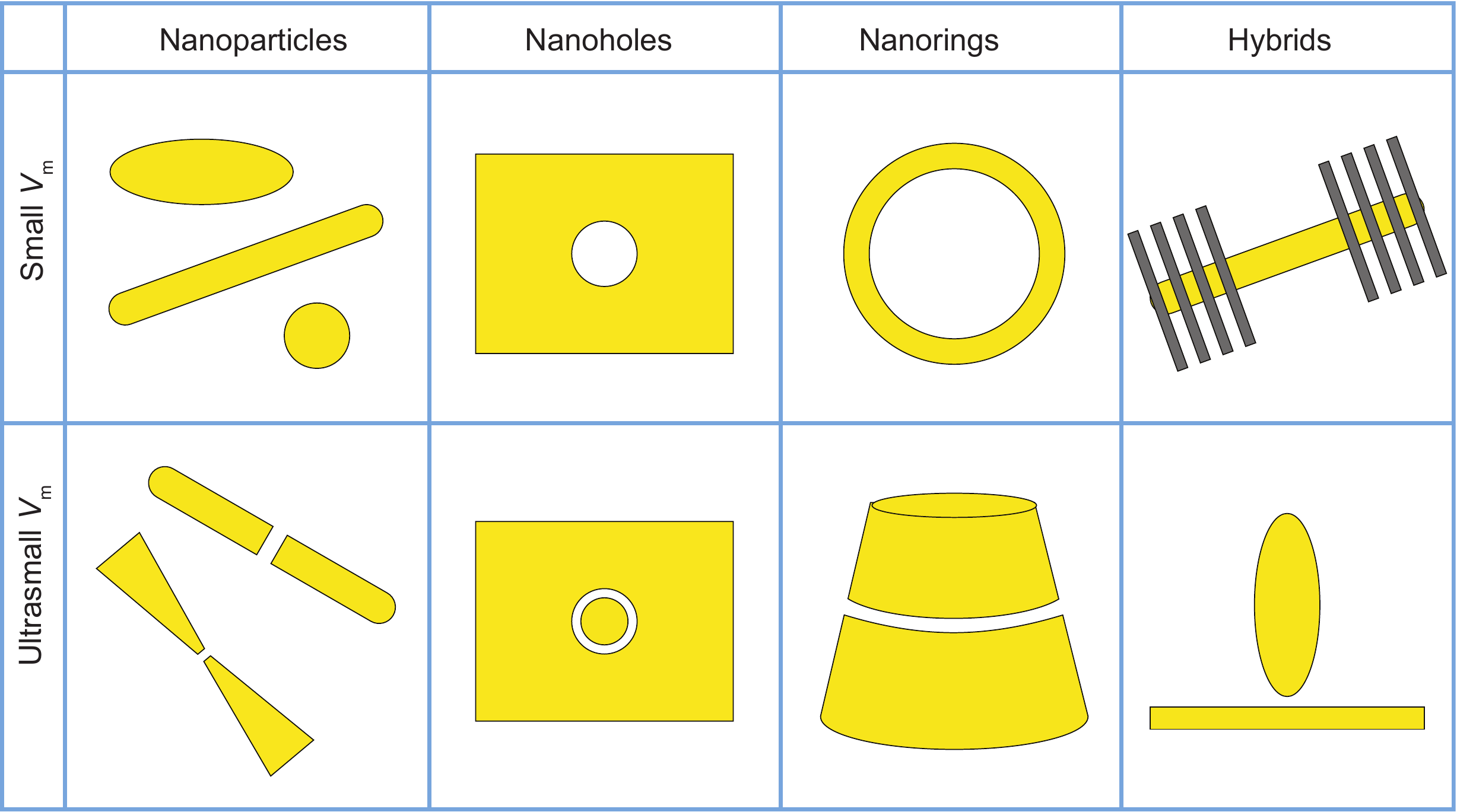}}
\caption{\label{nanoscale-resonators:antennas}
Classification of optical antennas according to their mode
volume $V_\mathrm{m}$ and confinement method, namely
nanoparticles,\cite{moskovits85,rogobete07a,mohammadi08b}
nanoholes and nanocavities,\cite{genet07,kroekenstoel09,maksymov10,bulu11}
nanorings\cite{aizpurua03,kuttge10} and hybrid
approaches.~\cite{le05,barth10,snapp10,eghlidi09,devilez10} 
Nanoholes and nanocavities may be understood as the complementary
structure of metal nanoparticles. Nanorings and nanodisks of deep
subwavelength dimensions should be considered as nanoparticles.
Otherwise they correspond to whispering gallery resonators.
Hybrid approaches range from the combination of nanoscale
structures with optical microcavities (upper panel) to the exploitation of
near-field effects with non-resonant geometries like a
metal film (bottom panel). Note that sorting based on the $Q$ factor,
as in Fig.~\ref{nanoscale-resonators:resonators}, would be less
meaningful.}
\end{figure*}

We based our discussion on a simplified antenna model,
which is nevertheless able to relate the main physical
magnitudes of a resonator with those of an antenna
and provide scaling laws for the figures of merit.
Moreover, we have found good agreement between
the outcome of our model and realistic antenna
designs~\citep{mohammadi08b}. These are
indicated as filled circles in the previous figures.
Although we based the analysis on metal nanoparticles,
we wish to point out that our expressions 
are in principle applicable to a larger class
of antennas and nanocavities,
since the geometrical factor $L$ is the only quantity that
depends on the specific design.
This is confirmed in Fig.~\ref{nanoscale-resonators:spontaneous-emission},
where we show that the parameters of a nanocavity
fit our model very well.

Inspired by Fig.~\ref{nanoscale-resonators:resonators},
we wish to conclude our analysis by presenting
in Fig.~\ref{nanoscale-resonators:antennas} a classification
of optical antennas according to their mode volume and confinement
method. In doing so we keep in mind that this field
is still making rapid progress.
Our attempt is thus to indicate which
approaches are consolidating
and how they may differ from conventional strategies
that have been used in the past century
to confine light at optical frequencies.
The striking differences with respect to
Fig.~\ref{nanoscale-resonators:resonators} must
not only be attributed
to the role of SPP resonances, but also
to the different level of theory involved in the resonator
design. In fact, while optical microcavities rely on
physical optics, nanoscale cavities owe their
properties to near-field optics, whose wealth of
effects may lead to unprecendented possibilities in the
resonator design.\cite{lukyanchuk10,halas11}

\section{Conclusions and outlook}

We investigated fluorescence enhancement by optical
antennas. Previous works indicated
that at optical wavelengths losses by real metals could quench light
emission. We established that this is not a fundamental constraint and
showed that the interaction can be improved by more than three orders
of magnitude without substantial quenching.\cite{rogobete07a}

We took advantage
of computational nano-optics to analyze the significant role that geometrical
details play in determining the antenna behavior.\cite{mohammadi08b,mohammadi10}
Moreover, we discussed the choice of different metals to enhance
emitters from the ultraviolet to the near-infrared spectral
range.\cite{mohammadi09a}
We would like to emphasize that these performances occur for
distances such that microscopic effects
can be safely neglected\cite{persson78,ford84,leung90}
and our design strategies are solely based
on electrodynamics considerations
like for radio-wave antennas.\cite{balanis05}

Optical antennas that strongly enhance the SE rate may
improve the quantum yield of weak emitters,
such as silicon nano-crystals,\cite{biteen05} molecules,\cite{kinkhabwala09}
nanotubes\cite{oconnell02} or diamond color centers,\cite{turukhin96}
and provide a handle on photophysical processes in general.\cite{nitzan81,mackowski08}
Furthermore, a larger decay rate permits a higher degree of
light emission, with immediate implications for single-photon
sources.\cite{lounis05,schietinger09}

An important theme of our research has been the enhancement
of light-matter interaction
towards levels pertaining to optical resonators.
To better understand the implications of these findings,
we derived figures of merit using
antenna theory\cite{hansen81} and compared
them to common resonator designs.\cite{vahala03,song05} Despite
absorption losses we found that antennas are promising
candidates for implementing the functionalities of 
optical resonators at the nanoscale.
Moreover, having a low
$Q$ factor, antennas do not suffer from the bandwidth
limitations that are common to high-finesse cavities.

Altogether, these settings hold great promise for interfacing
photons to a quantum 
system beyond the framework of cavity QED\cite{haroche89}
and urge further thorough theoretical and experimental investigations. 
These include studying the quantum optical phenomena that take place 
when an optical antenna mediates the interaction between photons and 
single quantum emitters in the full QED picture and beyond
continuous wave excitation.\cite{ridolfo10}
For instance, a number of proposals for quantum information science that
are based on cavity-assisted interactions could be explored
in this way.\cite{turchette95,vanloock06}

The ultrafast response of optical antennas, combined with their ability
to funnel light beyond the diffraction limit with a high
throughput,\cite{chen10}
has immediate implications for scanning implementations of time-resolved,
multidimensional and nonlinear
nanoscopies.\cite{sanchez99,mukamel99,guenther02,ichimura04,hartschuh08,abramavicius09}
Furthermore, combining ultrafast spectroscopy, field-enhanced
spectroscopy and quantum optics could push forward the possibility
of the coherent optical access of a quantum emitter above cryogenic
temperatures,\cite{brinks10,hildner11}
and monitor quantum coherence under conditions where dephasing
processes occur at very short time
scales.\cite{engel07,panitchayangkoon10}

The stringent requirements on photon management 
imposed by quantum-optical applications might turn out
to be extremely useful also for classical information 
processing to achieve, for instance, nonlinearities
at the single-photon level.\cite{chang07b}
In both cases we have to fight the mismatch
between light and nanoscale matter to attain strong and controllable
interactions; we need to process very small optical signals,
ideally down to single photons and possibly at very high rates. 
Thus, studying the physics and engineering of optical antennas may
also pave the way to the next generation of nanophotonics devices.\cite{miller89,miller09}

\section*{Acknowledgments}

M.A. wish to thank V. Sandoghdar for continuous support and advice.
He is also grateful to A. Mohammadi, X.-W. Chen, F. Kaminski and
L. Rogobete for the stimulating and fruitful collaboration.
This work was financed by ETH Zurich.

%%%%%%%%%%%%%%%%%%%%%%%%%%%%%%%%%%%%%%%%%%%%%%%%%%%%%%%%%%%%%%%%%%%%%
%% The appropriate \bibliographystyle and \bibliography commands
%% should be placed here.
%%%%%%%%%%%%%%%%%%%%%%%%%%%%%%%%%%%%%%%%%%%%%%%%%%%%%%%%%%%%%%%%%%%%%
\footnotesize{
\bibliographystyle{rsc}
%\bibliography{manuscript}

\providecommand*{\mcitethebibliography}{\thebibliography}
\csname @ifundefined\endcsname{endmcitethebibliography}
{\let\endmcitethebibliography\endthebibliography}{}

}

\end{document}